# Spatio-Temporal Coupled Mode Theory for Nonlocal Metasurfaces

Adam Overvig[1], Sander A. Mann[1], and Andrea Alù[1,2,*]

[1]Photonics Initiative, Advanced Science Research Center, City University of New York, New York, NY 10031, USA

[2]Physics Program, Graduate Center of the City University of New York, New York, NY 10016, USA

*Corresponding author: aalu@gc.cuny.edu

## Abstract

*Diffractive nonlocal metasurfaces have recently opened a broad range of exciting developments in nanophotonics research and applications, leveraging spatially extended - yet locally patterned - resonant modes to control light with new degrees of freedom. Their operation generalizes the response of periodic gratings, which conventionally resonate only for selected plane waves at specific combinations of incident angle and frequency (momentum-frequency selectivity). Nonlocal metasurfaces, empowered by quasi-bound states in the continuum (q-BICs), can be tailored to fully resonate for specific combinations of custom shaped wavefronts and frequencies (space-frequency selectivity). While conventional grating responses are elegantly captured by temporal coupled mode theory (TCMT), which is commonly employed to model and design grating responses across a wide application space, TCMT is not well equipped to capture the more sophisticated response observed in the nascent field of nonlocal metasurfaces, in which the eigen-states are arbitrarily shaped in space and hence not well confined in terms of momentum. Here, we introduce spatio-temporal coupled mode theory (STCMT), capable of elegantly capturing the key features of the resonant response of wavefront-shaping nonlocal metasurfaces. This framework can quantitatively guide nonlocal metasurface design, and is compatible with local metasurface frameworks, making it a powerful tool to rationally design and optimize a broad class of ultrathin optical components. We validate this STCMT framework against full-wave simulations of various nonlocal metasurfaces, demonstrating that this tool offers a powerful semi-analytical framework to understand and model the physics and functionality of these devices, without the need for computationally intense full-wave simulations. We also discuss how this model may shed physical insights into nonlocal phenomena in photonics and into the functionality of the resulting devices. As a relevant example, we showcase STCMT's flexibility by applying it to study and rapidly prototype nonlocal metasurfaces that spatially shape thermal emission.*

# I. Introduction

Over the years, coupled mode theory (CMT) has become a powerful tool to model and understand the scattering response of complex resonant electromagnetic structures [1]-[2]. Central to CMT is the phenomenological identification of: (i) the resonant modes supported by the system, and (ii) ports, or channels, that carry energy to and from the resonant scatterer. In turn, CMT provides an intuitive and accurate framework to study a wide range of resonant devices. In the spatial domain [3]-[5], it is commonly used to study waveguide modes and their coupling; while in the time domain temporal coupled mode theory (TCMT) [6]-[7] has found marked success in modelling spectral features such as Fano resonances and the construction of scattering matrices. This technique can be applied both to localized resonances [8]-[11] and to extended resonances supported by metasurfaces, photonic crystal slabs and subwavelength gratings [12]-[17]. However, in TCMT the spatial distribution of the mode is ignored: the modal amplitude $a$ is assumed to be function only of time. This feature in turn limits TCMT to studying the spectral features of the device, while detailed spectro-spatial information of the modes themselves is not captured. For this reason, TCMT is primarily effective at studying infinitely periodic grating structures, while finite and spatially varying gratings cannot be straightforwardly modeled.

The spectro-spatial properties of photonic resonant systems are important in various settings, particularly when the optical energy is correlated across distances larger than the wavelength, i.e., when nonlocality, also known as spatial dispersion, cannot be neglected. In turn, these systems exhibit strong dependence of the optical response on the momentum of light [18]-[19]. A sharp response in momentum space is referred to as *nonlocal* because the corresponding extent of the resonant modes in real space is necessarily broad. Nonlocality in the material permittivity has been studied in plasmonic systems for decades [20]-[23], associated with delocalized material responses that link the polarization field at a given point to the electromagnetic fields at neighboring points. More recently, epsilon-near-zero materials [24]-[26] and metamaterials [27]-[31] have been shown to support stronger nonlocal interactions with light. In dielectric thin film platforms, the dispersion of guided modes provides an engineerable mechanism to control even longer-range nonlocal phenomena [32]-[34]. In this context, nonlocal metasurfaces [35]-[37] have been gaining increasing attention for their capability of enhancing the control of light by tailoring it in momentum space, of particular interest for signal processing [36], image differentiation [37]-[39], spectral [40] and chiral sensing [41], augmented reality applications [42], and for mimicking the effects of free-space wave propagation within compact systems [43]-[45]. This class of devices employs subwavelength, transversely invariant structures, hence the spatial features of their extended resonant modes do not escape the near field of the device. As a result, when applied to infinitely periodic systems TCMT remains very effective, because it can lump their resonance into a spatially invariant modal parameter $a(t)$ [Fig. 1(a)].

Recent reviews [46], perspectives [47]-[49] and studies on the fundamental role of nonlocality in optics [50] suggest that embracing nonlocality is a budding area of inquiry and a compelling direction for the maturing field of metasurfaces. In particular, it has been realized that engineered nonlocalities can be combined with locally varying features to enable *diffractive nonlocal metasurfaces* [48]. These ultrathin optical devices support highly selective resonant responses combined with tailored spatial variations, which are capable of patterning the optical wavefront engaging the resonance [51]-[60]. Leveraging the physics of quasi-bound states in the continuum (q-BICs) [61], these devices support long-lived, spatially-extended resonant states that couple to free-space with amplitude, phase and polarization properties spatially tailored across the aperture through symmetry-controlled perturbations. By spatially varying the perturbations with a uniform phase gradient, for instance, light can be anomalously diffracted with near-unity efficiency and circular polarization selectivity [54]. Their functionality is not limited to beam steering: by nonperiodically tailoring the local perturbations across the device, the resonant response can efficiently engage arbitrarily tailored optical wavefronts of choice. As a remarkable example, a nonlocal metalens with hyperbolic phase distribution across the aperture can selectively engage only fields originating from its focal spot, while it remains transparent for other wavefront shapes [55]. This spatially selective response, for which the metasurface is only resonant for specific incoming wavefront shapes that match its local pattern of perturbations, generalizes the angular selectivity well-known for resonant gratings, which is limited to plane waves. Since their response is spectro-spatial in a nontrivial way, however, it cannot be captured by TCMT. Rather, a new theoretical framework featuring a spatially varying modal parameter $a(x,t)$ is required [Fig. 1(b)]. Figure 1(c) depicts an example nondiffractive nonlocal metasurface (based on Ref. [54]), whose meta-unit cells are subwavelength and transversely invariant, hence compatible with a TCMT description. Figure 1(d) on the contrary depicts an example diffractive nonlocal metasurface based on the same meta-unit cells, but with a spatially varying orientation angle (from left to right), incompatible with a TCMT description and requiring a generalized CMT model.

To address this need, here we introduce a general spatio-temporal coupled mode theory (STCMT) that extends TCMT to capture spatially varying resonant modes. We show excellent agreement with full-wave simulations of spatially selective diffractive nonlocal metasurfaces [55], demonstrating the validity and relevance of this model. Our theory introduces a *nonlocality length*, which quantifies the degree of nonlocality of a given device, and it establishes how this parameter directly controls the degree of spatial selectivity. We clarify two distinct regimes of operation for nonlocal photonic devices, governed by the numerical aperture of the anomalous diffraction encoded into the device, and we demonstrate the relation between spatial selectivity and the supported eigenmodes. As such, STCMT elegantly captures the fundamental working mechanisms of spatially varying resonant photonic systems, and it provides physical insights into the nature of nonlocality in this emerging class of novel devices, enabling rational designs and rapid, computationally efficient analysis of next-generation photonic systems.

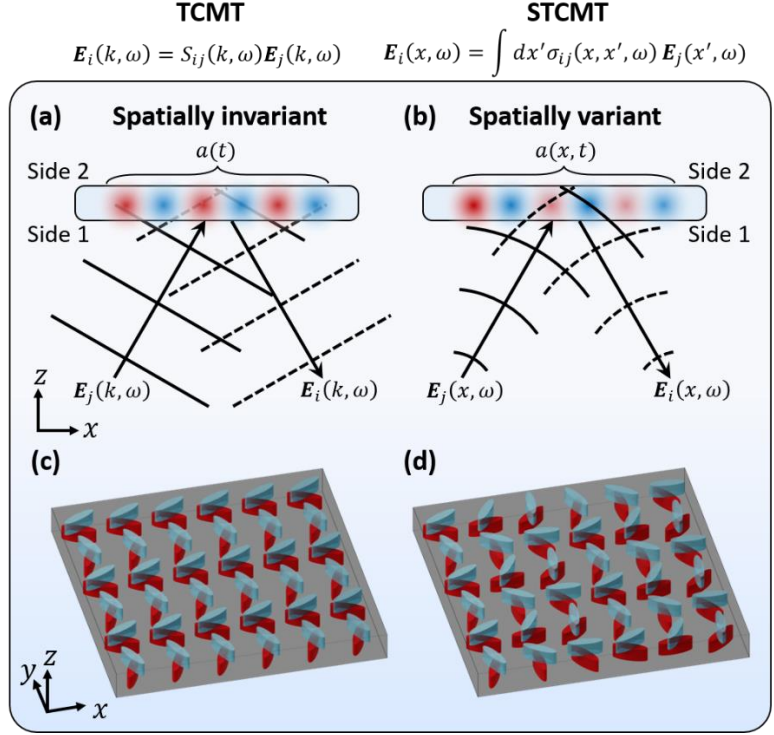

**Figure 1. Coupled mode theories.** (a) TCMT is effective at modeling systems in which the complex modal amplitude $a(t)$ is spatially invariant, wherein the scattered field $\mathbf{E}_j$ is the product of the scattering matrix $S_{ij}$ and the incident field $\mathbf{E}_i$. (b) STCMT is necessary when the complex modal amplitude $a(x,t)$ is spatially varying, wherein the scattered field $\mathbf{E}_j$ is the convolution of the scattering kernel $\sigma_{ij}$ and the incident field $\mathbf{E}_i$. (c) Example nondiffractive nonlocal metasurface with spatially invariant meta-unit cells. (d) Example diffractive nonlocal metasurface with spatially varying meta-unit cells, varying from left to right.

## II. Spatio-temporal coupled mode theory

In the following, we develop STCMT by writing down the appropriate dynamical equations, determining the constraints stemming from time-reversal invariance, reciprocity and energy conservation, and then computing the scattering from a general nonlocal device by appropriately specifying its parameters.

### II.A Dynamical equations

We obtain the spatio-temporal dynamical equations by appropriately manipulating and tailoring standard TCMT equations [6]. Under an $e^{-i\omega t}$ time convention, and focusing on a single resonance, these equations are

$$\frac{da(t)}{dt} = -i\Omega a(t) + \langle \kappa^* | s_+ \rangle \quad (1)$$

$$|s_-(t)\rangle = C|s_+(t)\rangle + a(t)|d\rangle. \quad (2)$$

Equation (1) describes the dynamics of the complex modal amplitude $a(t)$ excited by an incoming wave $|s_+(t)\rangle$, where $|a(t)|^2$ is normalized as the energy stored in the q-BIC per unit area at time $t$ and $\langle s_+ | s_+ \rangle$ is the incident intensity. The in-coupling vector $|\kappa\rangle$ contains the coupling coefficients to the mode from free-space excitation, and $\Omega = \omega_0 - i\gamma$ is a complex frequency, with $\omega_0$ being the (resonant) modal frequency and $\gamma$ being the total scattering rate, which we take to be purely radiative throughout. Extension to include absorption loss is straightforward. Equation (2) describes the dynamics of the outgoing (scattered) waves $|s_-\rangle$ (also normalized such that $\langle s_- | s_- \rangle$ is the outgoing intensity) as the interference of the "background" scattering, described by the matrix $C$, and the mode leaking to free space, described by the out-coupling vector $|d\rangle$ (see **Supplementary Section S1** for additional background). Notably, the first-order nature of the temporal dynamics in conventional TCMT is rooted in Maxwell's equations, regardless of the modal properties [1].

In contrast, the dynamical equations for a device with space-varying modal properties must inherently depend on the nature of the mode itself. In other words, we need at the outset to know the form of the dispersion of the mode in order to capture its spatial properties. We expect, for instance, the phenomenology of a travelling wave to differ from the one of a standing wave, and hence we expect distinct dynamical equations. The most general approach to capture this dependence is a Taylor expansion for the resonance frequency in terms of its momentum $k$ dependence:

$$\Omega(k) = \left(\omega_0 + ck + \frac{b}{2}k^2 + ...\right) + i\left(\gamma_0 + \gamma_1 k + \frac{\gamma_2}{2}k^2 ...\right). \quad (3)$$

For instance, a traveling wave is captured by the linear term with speed $c$, while a standing wave with parabolic dispersion, such as near a band edge, is captured by the coefficient $b$. As detailed in **Supplementary Section S2**, transforming Eqns. (1) and (2) into the frequency domain, including its momentum dependence and the expansion (3), and then transforming to the space-time domain, yields the dynamical equations (up to second order)

$$\frac{da(x,t)}{dt} + i(\omega_0 + i\gamma_0)a(x,t) - (c+i\gamma_1)\frac{da(x,t)}{dx} + i(b+i\gamma_2)\frac{d^2a(x,t)}{dx^2} = \int dx' \langle \kappa(x,x')|s_+(x')\rangle \quad (4)$$

$$|s_-(x)\rangle = \int dx' \left(C(x,x')|s_+(x')\rangle + a(x)|d(x,x')\rangle\right). \quad (5)$$

Equations (4) and (5) are the dynamical equations for STCMT, including spatial dispersion in both the *coupling* properties (convolution terms on the right-hand sides) and the *modal* properties (spatial derivative terms on the left-hand sides). We note that this is can be readily extended to higher order expansions if needed, but in the most common scenarios the coefficients in Eqn. (4) are sufficient – often even some of the lower order expansion coefficients may be neglected. For instance, a pure traveling wave is characterized by a first-order spatial derivative, with Eqn. (4) simplifying to a form akin to the one-way wave equation; a standing wave at a band edge, modeled parabolically, is characterized by a second-order spatial derivative, with Eqn. (4) simplifying to an effective mass, non-Hermitian Schrödinger equation. The Taylor expansion coefficients $c, b$ therefore quantify the spatial dispersion of two overlapping modes of different type, each of which can be captured by STCMT. Similarly, we see that the radiative decay profile, captured by the coefficient $\gamma_{1,2}$, also describes a form of nonlocal effect, with or without the presence of $c, b$.

Next, we specialize our general STCMT to capture the novel functionalities demonstrated by recent diffractive nonlocal metasurfaces, as in Refs. [51]-[58]. These devices leverage q-BIC modes, whose Q-factors are sufficiently large such that the background varies slowly compared to the resonant features [6]. In their infinitely periodic implementation, they abide a TCMT description with $c = \gamma_1 = \gamma_2 = 0$ near normal incidence. In other words, they are well captured by a parabolic band with a Q-factor that is not a function of the incident angle [62]. Their design assumes that the scattering of a q-BIC can be spatially tailored by locally changing a small symmetry-breaking perturbation, and that distant perturbations do not affect the local polarization or phase [52]. Hence, the nonlocality of these devices is assumed to be purely *modal,* or captured by the spatial derivatives in the left-hand side, while the *coupling* to and from the mode is spatially instantaneous (local):

$$\begin{aligned} C(x,x') &\propto \delta(x-x') \\ |\kappa(x,x')\rangle &\propto \delta(x-x') \\ |d(x,x')\rangle &\propto \delta(x-x') \end{aligned} \quad (6)$$

Together, these assumptions yield the dynamical equations

$$\frac{da(x,t)}{dt} + i(\omega_0 + i\gamma)a(x,t) + ib\frac{d^2 a(x,t)}{dx^2} = \langle \kappa(x) | s_+(x,t) \rangle, \quad (7)$$

$$|s_-(x,t)\rangle = C(x)|s_+(x,t)\rangle + a(x,t)|d(x)\rangle, \quad (8)$$

where we simplify the notation to $\gamma = \gamma_0$. **Supplementary Section S2b** derives the same set of equations using the leading order terms of a cosine series expansion.

## II.B Scattering Matrix, Propagator and Green's Function

In TCMT, the scattering matrix is obtained by assuming time-harmonic solutions, solving Eqn. (1) for $a(t)$, and then inserting the result into Eqn. (2) to yield [6]

$$|s_-(\omega)\rangle = S(\omega)|s_+(\omega)\rangle \quad (9)$$

$$S(\omega) = C(\omega) + \frac{|d\rangle\langle k^*|}{i(\Omega - \omega)}. \tag{10}$$

In comparison, in STCMT we must solve Eqn. (7) before inserting it into Eqn. (8). To do so, we recognize that, when no source is present, we obtain a Schrödinger equation with complex potential $V = \omega_0 + i\gamma$, recognizable by the rearrangement:

$$i\frac{da(x,t)}{dt} = -\frac{b}{2}\frac{d^2 a(x,t)}{dx^2} + Va(x,t). \tag{11}$$

Compared to the conventional Schrödinger equation, we have $\hbar = 1$ and $b = 1/m$, consistent with the expectation that the band curvature plays the role of an inverse effective mass. We also note that non-Hermiticity (a complex valued potential) is expected of an open or damped system.

In absence of the potential $V$, we have the equation for a free particle, for which the solution is described by the propagator $K$ and the initial condition $a(x,t=0) = a_0(x)$:

$$a(x,t) = \int K(x,x',t) a_0(x') dx' \tag{12}$$

$$K(x,x',t) = \frac{1}{\sqrt{2\pi i b t}} e^{i(x-x')^2 / 2bt}. \tag{13}$$

The addition of a spatially constant $V$ adds a phase factor independent of $x$. Hence, in our case the propagator is

$$K(x,x',t) = \frac{1}{\sqrt{2\pi i b t}} e^{i(x-x')^2 / 2bt} e^{-i\omega_0 t} e^{-\gamma t}. \tag{14}$$

In addition to solving the homogenous equation, the propagator yields the Green's function $G_t$ for the inhomogeneous equation (7):

$$G_t(x,x',t) = \frac{1}{2\gamma} \Theta(t) K(x,x',t), \tag{15}$$

where $\Theta(t)$ is the Heaviside step function. Using these functions, we may apply physical constraints such as conservation of energy (which takes the form of a continuity equation), time-reversal invariance and reciprocity (see Section II.C and **Supplementary Section S4**).

In practice, we are interested in the response to monochromatic waves [$\exp(-i\omega t)$], i.e., we seek the solution to the scattering problem in the space-frequency domain. In this case, Eqn. (7) becomes

$$i(\omega_0 - \omega + i\gamma)a(x,\omega) + ib\frac{d^2 a(x,\omega)}{dx^2} = \langle \kappa(x) | s_+(x,\omega) \rangle, \tag{16}$$

and we seek the Green's function satisfying

$$i(\omega_0 - \omega + i\gamma)G(x,x',\omega) + ib\frac{d^2 G(x,x',\omega)}{dx^2} = \delta(x-x'), \tag{17}$$

which yields the modal amplitude

$$a(x,\omega) = \int dx' G(x,x',\omega) \langle \kappa(x') | s_+(x',\omega) \rangle \tag{18}$$

Insertion of Eqn. (18) into the space-frequency form of Eqn. (8) yields the scattering equation

$$|s_-(x,\omega)\rangle = \int dx' \sigma(x,x',\omega) |s_+(x',\omega)\rangle \tag{19}$$

$$\sigma(x,x',\omega) = C(x,\omega)\delta(x-x') + G(x,x',\omega)|d(x)\rangle\langle \kappa^*(x')|, \tag{20}$$

where the scattering kernel $\sigma$ correlates input positions $x'$ with output positions $x$, and where the closed form of the Green's function is

$$G(x,x',\omega) = -i\frac{1}{b}\xi(\omega)\exp\left(-\frac{|x-x'|}{\xi(\omega)}\right), \tag{21}$$

parameterized by the complex coefficient

$$\xi(\omega) = \sqrt{\frac{ib/2}{\gamma + i(\omega_0 - \omega)}}. \tag{22}$$

This quantity was originally introduced to capture the spatial coherence of thermal metasurfaces, derived with alternative methods [56]. Its real part takes an especially simple form at the band edge

$$\xi_0 = \text{Re}\left[\xi(\omega_0)\right] = \sqrt{b/\gamma} = \sqrt{b\tau_r}, \tag{23}$$

which we call the *nonlocality length*. This length is the characteristic distance that light travels in-plane before scattering out when engaging with the band-edge mode; in other words, it is a quantitative measure of the nonlocality of the metasurface at the band-edge frequency. Since nonlocality is equivalently defined as a dependence on the incident wavevector, no optical interface is truly local. However, as a rule of thumb, if the nonlocality length is larger than the wavelength in the surrounding media, the resulting device may be considered nonlocal, since its nonlocality cannot be neglected (see **Supplementary Section S3** for a comparison of a nonlocal metasurface to a bare interface between two media).

### II.C Physical Constraints

Before proceeding with the application of STCMT to practical metasurface devices, we must determine the relevant physical constraints to our system, including time-reversal invariance, reciprocity and conservation of energy. In TCMT, these constraints impose the well-known requirements [6]

$$\langle d|d \rangle = 2\gamma \tag{24}$$

$$|\kappa\rangle = |d\rangle \tag{25}$$

$$C|d^*\rangle = -|d\rangle. \tag{26}$$

These results hold, for instance, in infinitely periodic devices supporting q-BICs (**see Supplementary Section S1**). As detailed in **Supplementary Section S4**, we find that in our STCMT all these conditions hold *locally*, i.e.,

$$\langle d(x)|d(x)\rangle = 2\gamma \tag{27}$$

$$|\kappa(x)\rangle = |d(x)\rangle \tag{28}$$

$$C(x)|d^*(x)\rangle = -|d(x)\rangle. \tag{29}$$

We stress that this result is not obvious *a priori* and, indeed, it does not hold for the general case described by Eqns. (4) and (5). Future work may explore systems with nonlocal coupling (i.e., devices for which Eqns. (6) do not hold), in which case Eqns. (28) and (29) are no longer valid.

The significance of this result for modelling diffractive nonlocal metasurfaces should not go underappreciated. This property stems from the *shift-invariance* of the Green's function in Eqn. (21), meaning that the spatial dependence of the nonlocal scattering is wholly contained in the coupling coefficient $|d(x)\rangle$ in Eqn. (20) [where we note $\langle \kappa^*(x')| = \langle d^*(x')|$ per Eqn. (28)]. The assumption that the modal properties $b, \gamma, \omega_0$ are space-invariant, i.e., that the local perturbations tailoring the spatial response of the metasurface do not affect them, and that the coupling is spatially instantaneous [Eqn. (6)], confer a major simplification to model this class of systems and enables the introduction of a semi-analytical model to describe them. In particular, each unit cell of a nonlocal metasurface may be modeled as if it were in an infinite array, allowing easy correlation between geometric and phenomenological degrees of freedom, compatible with TCMT in the momentum-frequency domain. Then, a given space-varying nonlocal metasurface may be modelled by simply specifying a spatial profile of the coupling coefficients, wherein the correlations across the surface are captured in STCMT by the shift-invariant Green's function. This design flow is compatible with the foundational design principles of *local* metasurfaces [63], wherein a library of pre-computed optical scatterers may be arrayed across the surface as if they operate independently of their nearest neighbors. This ubiquitous design approach is sometimes called the "local phase approximation". Yet, we find that, when Eqn. (6) holds, the same design principles are applicable even in the much more complex scenario discussed in this work, considering arbitrarily *nonlocal* responses due to a guided mode.

### II.D Eigenwaves

STCMT enables us to rigorously define characteristic 'eigenwaves' for nonlocal metasurfaces with arbitrarily varying spatial profiles. An eigenwave can be defined as a spatially varying wavefront that a nonlocal metasurface is selective for: maximal reflectance is achieved when the eigenwave is incident, and the reflected wave is the eigenwave's time-reversed copy (preserving coordinate system handedness, in contrast to specular reflection) [55]. Using $\rho = \sigma_{11}$ as the reflection kernel for light incident from side 1, the eigenwave from side 1 is solution to the relation

$$E_{eig}^*(x,\omega) = \int dx' \rho(x,x',\omega) E_{eig}(x',\omega), \qquad (30)$$

or equivalently

$$E_{eig}(x,\omega) = \int dx' \rho^*(x,x',\omega) E_{eig}^*(x',\omega). \qquad (31)$$

Combing Eqns. (30) and (31) we find

$$E_{eig}(x,\omega) = \int dx' \int dx'' \rho^*(x,x',\omega) \rho(x',x'',\omega) E_{eig}(x'',\omega), \qquad (32)$$

suggesting that the eigenwave of interest is the eigenvector of

$$R_{eig} E_{eig}(x,\omega) = \int dx' \int dx'' \rho^*(x,x',\omega) \rho(x',x'',\omega) E_{eig}(x'',\omega) \qquad (33)$$

that has the largest eigenvalue $R_{eig}$, i.e., the eigenvector with maximal reflectance. Equation (33) implies the existence of higher-order eigenwaves that reflect to their time-reversed copy but with non-maximal reflectance. It also suggests the existence of eigenwaves for frequencies other than the band-edge frequency $\omega_0$.

### III. Diffractive Nonlocal Metasurfaces

In this section, we apply the introduced STCMT to illustrative examples, confirming that our analytical framework reproduces the results of full-wave simulations, quantitatively captures the underlying physics and enables rational designs for relevant functionalities. Schematically depicted in Figure 2(a), we explore systems based on Refs. [54] and [55], composed of four elliptical inclusions controlling the scattering (polarization angles $\phi$ and $\theta$) at the top and bottom interfaces simultaneously and independently [Fig. 2(b)]. Figure 2(c,d) show the calculated reflectance and reflected phase for an example meta-unit based on this system, computed using the finite difference time domain (FDTD) method. Figures 2(e,f) depict the TCMT modelling of the infinitely periodic meta-unit (developed in **Supplementary Section S1**), with (in units such that $c=1$) the band edge frequency $\omega_0 = 2\pi / \lambda_{be}$ with $\lambda_{be} = 1.6104 \mu m$, the band curvature $b = 0.021 \mu m$, and the lifetime $\tau_r = 250 \mu m$.

In particular, we use the chiral q-BIC system in Refs. [54] and [55], in which the response was shown to: (i) have zero background reflection, (ii) exclusively engage circularly polarized light of one handedness, and (iii) decay with a geometric phase of $2\alpha$ for a meta-unit characterized by in-plane orientation angle $\alpha$. After judiciously choosing the geometrical parameters defining the ellipses, we may use a scalar basis (i.e., ignore the opposite handedness of circularly polarized light, which does not engage the resonance), and up to a constant reference phase we have (see **Supplementary Section S1**)

$$|d\rangle = i\sqrt{\gamma} \begin{bmatrix} \exp[i2\alpha] \\ -i\exp[-i2\alpha] \end{bmatrix}, \tag{34}$$

where the first element is the coefficient scattering to side 1 and the second element is the coefficient scattering to side 2. In diffractive nonlocal metasurfaces, we allow $\alpha$ to vary spatially:

$$|d(x)\rangle = i\sqrt{\gamma} \begin{bmatrix} \exp[i2\alpha(x)] \\ -i\exp[-i2\alpha(x)] \end{bmatrix}. \tag{35}$$

For a system based on the meta-units in Fig. 2 (and Refs. [54]-[55]), this yields [from Eqns. (20) and (21)]

$$\sigma(x,x',\omega)=\delta(x-x')\begin{bmatrix} 0 & -i \\ -i & 0 \end{bmatrix}+i\frac{\gamma}{b}\xi(\omega)\exp\left(-\frac{|x-x'|}{\xi(\omega)}\right)\begin{bmatrix} e^{i2\alpha(x)}e^{i2\alpha(x')} & -ie^{i2\alpha(x)}e^{-i2\alpha(x')} \\ -ie^{-i2\alpha(x)}e^{i2\alpha(x')} & -e^{-i2\alpha(x)}e^{-i2\alpha(x')} \end{bmatrix}. \quad (36)$$

To improve clarity, in what follows we detail the behavior for light incident from side 1, in which case we may separate the complex scattering kernel into a complex reflection kernel $\rho=\sigma_{11}$ and a complex transmission kernel $\tau=\sigma_{21}$:

$$\rho(x,x',\omega)=i\frac{\gamma}{b}\xi(\omega)\exp\left(-\frac{|x-x'|}{\xi(\omega)}\right)e^{i2\alpha(x)}e^{i2\alpha(x')}$$
$$\tau(x,x',\omega)=-i\delta(x-x')+\frac{\gamma}{b}\xi(\omega)\exp\left(-\frac{|x-x'|}{\xi(\omega)}\right)e^{-i2\alpha(x)}e^{i2\alpha(x')} \quad (37)$$

The other kernels $\sigma_{12}$ and $\sigma_{22}$ are treated equivalently, but omitted here for brevity. Figures 2(g,h) show the magnitude and phase of $\sigma_{11}$ for the meta-unit corresponding to Figs. 2(e,f). Note that in this case the phase angle $\alpha(x)=\alpha$ is constant, and there is only one diffraction order on each side, meaning that this metasurface can be handled with conventional TCMT. In the following examples, however, the angle is spatially varied, and the use of regular TCMT is not possible. We will then demonstrate that STCMT readily offers excellent agreement with numerical simulations and captures the underlying physics.

### III.A Linear phase gradient: analytical model

In this sub-section, we focus on nonlocal metasurfaces with a linear phase gradient. In particular, we consider a gradient in the geometric angle following

$$2\alpha(x)=-k_G x. \quad (38).$$

In this case, the nonlocal kernels take the form

$$\rho(x,x',\omega)=i\frac{\gamma}{b}\xi(\omega)\exp\left(-\frac{|x-x'|}{\xi(\omega)}\right)e^{-ik_G x}e^{-ik_G x'}$$
$$\tau(x,x',\omega)=-i\delta(x-x')+\frac{\gamma}{b}\xi(\omega)\exp\left(-\frac{|x-x'|}{\xi(\omega)}\right)e^{ik_G x}e^{-ik_G x} \quad (39)$$

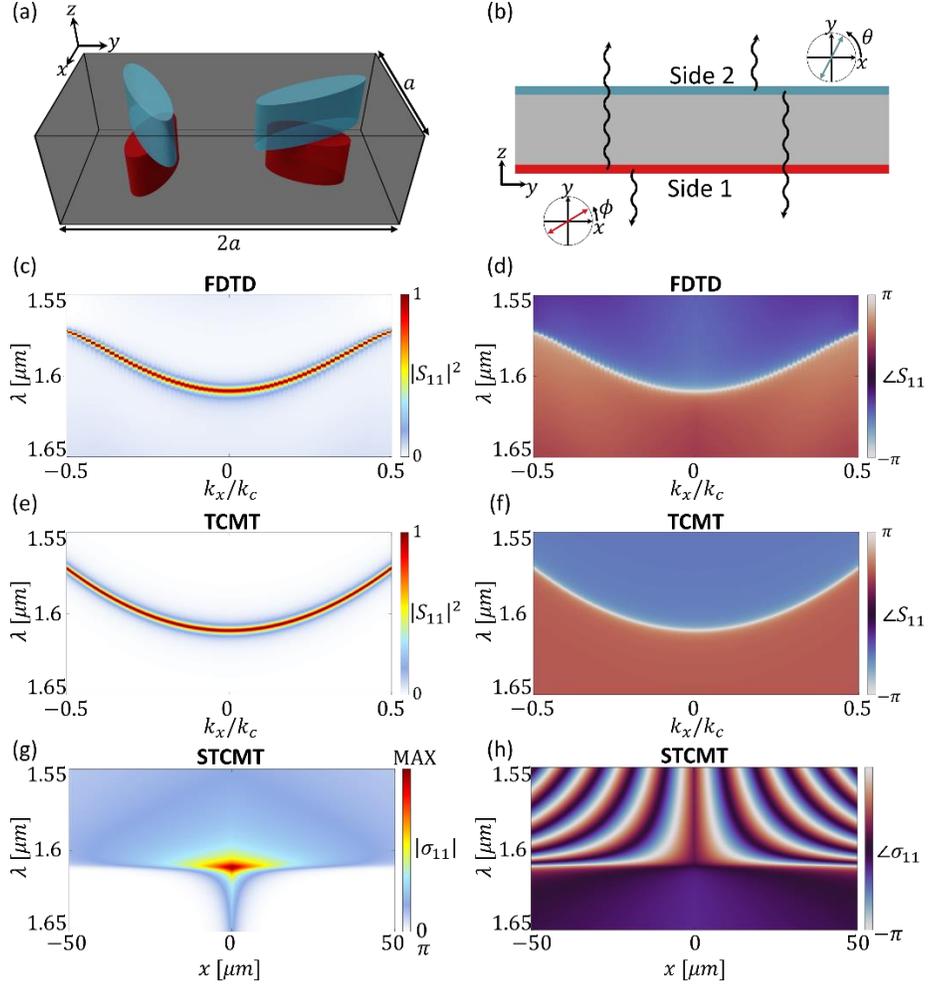

**Figure 2.** (a) Prototypical unit cell for a diffractive nonlocal metasurface, containing four elliptical inclusions. The lattice parameter $a$ satisfies $2a < \lambda_0$, eliminating all but 1 diffraction order on each side. (b) Schematic of the coupled mode theory approximation of the scattering in (a), wherein independent scattering events occur at the interfaces, each of which may generally couple to both side 1 (down) and side 2 (up). Reflectance (c) and reflection phase (d) in the momentum-frequency response of an example meta-unit, computed by Finite Difference Time Domain (FDTD). Reflectance (e) and reflection phase (f) in the momentum-frequency response of an example meta-unit, modeled using TCMT. Amplitude (g) and phase (h) of $\sigma_{11}$ in the space-frequency domain of the same meta-unit, a key object for modeling this system using STCMT.

When $k_G = 0$, we recover the nonlocal reflection kernel without phase gradient, which we denote with the subscript $0$:

$$\rho_0(x,x',\omega)=i\frac{\gamma}{b}\xi(\omega)\exp\left(-\frac{|x-x'|}{\xi(\omega)}\right)$$
$$\tau_0(x,x',\omega)=-i\delta(x-x')+\frac{\gamma}{b}\xi(\omega)\exp\left(-\frac{|x-x'|}{\xi(\omega)}\right)$$
(40)

Since a phase gradient provides a spatially constant momentum kick, it is natural to study these devices in momentum space. In particular, we seek the scattering matrix

$$S(k,k',\omega)=\begin{bmatrix} S_{11}(k,k',\omega) & S_{12}(k,k',\omega) \\ S_{21}(k,k',\omega) & S_{22}(k,k',\omega) \end{bmatrix},$$
(41)

where $k'$ and $k$ are the basis wavevectors of the incoming and outgoing wavevectors, respectively, and the indices refer to the top and bottom sides of the metasurface. We then need to relate this scattering matrix to the scattering kernel $\sigma$, which, as shown in **Supplementary Section S5**, is the mixed Fourier transform

$$S_{ij}(k,k',\omega)=\int dx\int dx'\sigma_{ij}(x,x',\omega)\exp(ik'x')\exp(-ikx).$$
(42)

Using this relation, the scattering components for a metasurface without phase gradient are

$$S_{11}^0(k,k',\omega)=\frac{-\delta(k-k')}{1-i\left(\omega-\omega_0-\frac{b}{2}k^2\right)\tau_r}$$

$$S_{21}^0(k,k',\omega)=\delta(k-k')\left[-i+\frac{i}{1-i\left(\omega-\omega_0-\frac{b}{2}k^2\right)\tau_r}\right].$$
(43)

$$S_{22}^0(k,k',\omega)=S_{11}^0(k,k',\omega)$$
$$S_{12}^0(k,k',\omega)=S_{21}^0(k,k',\omega)$$

Notably, this result is consistent with TCMT, with the explicit addition of a Dirac $\delta$ term enforcing scattering only when $k=k'$ (conservation of momentum in a specular process).

Meanwhile, the scenario with nonzero $k_G$ in terms of Eqn. (40) yields

$$\rho(x,x',\omega)=\rho_0(x-x',\omega)e^{-ik_Gx}e^{-ik_Gx'}$$
$$\tau(x,x',\omega)=\tau_0(x-x',\omega)e^{ik_Gx}e^{-ik_Gx'}$$
(44)

which implies, due to the basic properties of the Fourier transform

$$\begin{aligned}S_{11}(k,k',\omega)&=S_{11}^0(k+k_G,k'-k_G,\omega)\\ S_{21}(k,k',\omega)&=S_{21}^0(k-k_G,k'-k_G,\omega)\\ S_{22}(k,k',\omega)&=S_{11}^0(k-k_G,k'+k_G,\omega)\\ S_{12}(k,k',\omega)&=S_{12}^0(k+k_G,k'+k_G,\omega)\end{aligned} \quad (45)$$

The impact of a nonlocal phase gradient is to shift the scattering matrices in *k*-space, in both the *input* and *output* spaces. In particular, in the transmission elements [ $S_{12}(k,k',\omega)$ and $S_{21}(k,k',\omega)$ ] the shift in input and output momentum is identical, with $+k_G$ when light is incident from side 2 and $-k_G$ when incident from side 1. In contrast, in the reflection elements [ $S_{11}(k,k',\omega)$ and $S_{22}(k,k',\omega)$ ] the shift in input and output momentum is equal and opposite, producing anomalous reflection. Since we have nonzero scattering only when the Dirac delta function's argument is 0, from side 1 we see that reflection occurs only when $k=k'+2k_G$, and from side 2 only when $k=k'-2k_G$. Hence, our results confirm that the anomalous reflection is associated with the $m=\pm 2$ diffraction orders, and the direction of momentum kick imparted by the nonlocal metasurface depends on which side light is incident from.

### III.B Linear phase gradient: numerical simulations

We now validate our analytical model numerically for $k_G=2\pi/W$ and $W=6.4\mu m$, focusing here on the band-edge frequency. **Supplementary Section S6** extends the following analysis to a representative frequency off the band edge. To aid in this, it is useful to cast STCMT in matrix form, detailed in **Supplementary Section S7**. In this case, the scattering kernels are discrete, and we write them as P and T (capital $\rho$ and $\tau$). We write the input electric field as a column vector $\mathbf{E}_{in}$, where each entry is the electric field at a different discretized position $x$, and we compute the reflected $\mathbf{E}_r$ and transmitted $\mathbf{E}_t$ fields with

$$\begin{aligned}\mathbf{E}_r &= \mathrm{P}\mathbf{E}_{in}\\ \mathbf{E}_t &= \mathrm{T}\mathbf{E}_{in}\end{aligned}. \tag{46}$$

The main diagonal of P and T correspond to $x = x'$, i.e., the local response, while the off-diagonal components contain the correlations across distant positions, i.e., the nonlocal response. Meanwhile, the scattering matrix defined in Eqn. (41) also becomes discrete, and relation (42) may be conveniently computed using the Fast Fourier Transform method. Finally, given the finite nature, boundary conditions must be specified; **Supplementary Section S8** details both periodic and radiative boundary conditions. **Supplementary Section S9** describes the extension of this procedure to 2D metasurfaces.

Using this discretized form of STCMT, Figure 3 compares the scenario without a phase gradient [Figs. 3(a-d)] to the case with a phase gradient [Figs. 3(e-h)]. We emphasize that, while the former case may be straightforwardly treated with conventional TCMT, the latter may not. Figure 3(a) schematically shows resonant reflection at the band-edge frequency, $\omega = \omega_0$, depicting specular reflection at normal incidence when incident from either side. Figure 3(b) shows the reflectance profile as a function of wavelength and incident momentum from side 1, showing a parabolic band profile typical of a q-BIC. Figure 3(c) shows the corresponding amplitude and phase of the nonlocal reflection and transmission matrices from side 1, P and T (note that T is identical to P except along the main diagonal, where its magnitude is much higher). Finally, Fig. 3(d) shows the scattering matrix elements for this case, confirming that specular transmission arises except at normal incidence from either side, wherein specular reflection occurs. In order to achieve sufficient resolution in k-space, Fig. 3(d) was computed using the mixed Fast Fourier Transform of a version of the plot in Fig. 3(c) including 10 metasurface periods.

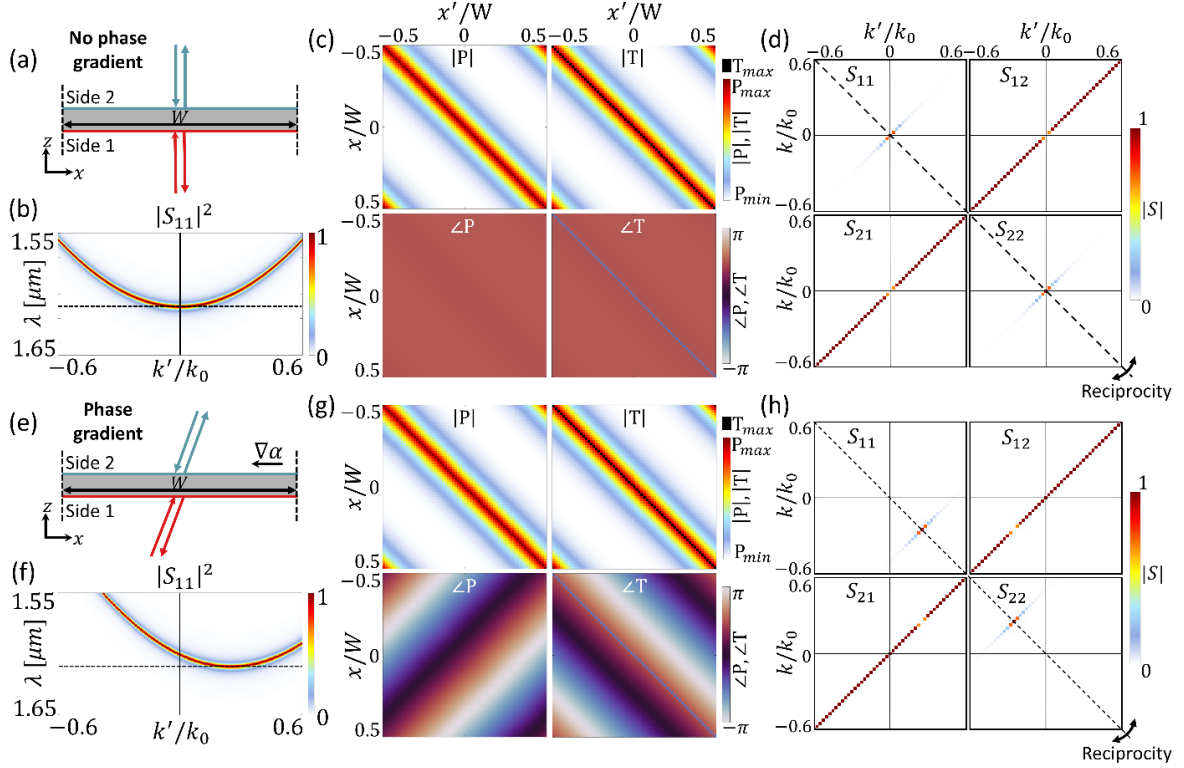

**Figure 3. Analysis of phase gradient devices.** (a) Schematic of the resonant response of a nonlocal phase gradient device with $k_G = 0$ at the band-edge frequency. (b) Reflectance as a function of incident momentum $k'$ for the device in (a), where the dashed line indicates the band-edge frequency. (c) Nonlocal kernel matrices for the device in (a) at the band-edge frequency. (d) Scattering matrix for the device in (a), showing unity specular reflectance only at normal incidence. (e-h) Same as in (a-d) but for a device with $k_G = 2\pi/W$ operated at the band-edge frequency.

In contrast, in the presence of a phase gradient the device retroreflects light at the resonance frequency, as shown in Fig. 3(c). From side 1, the band edge mode is shifted in k-space by $k_G$, meaning that the resonance only occurs for $k' = k_G$. In particular

$$\omega(k') = \omega_0 + \frac{b}{2}(k' - k_G)^2. \tag{47}$$

This shift, as well as the anomalous reflection, is encoded in P and T within the phase information [Fig. 3(g)], and the scattering matrix at the band-edge frequency is shown in Fig. 3(h), confirming that the resonant mode occurs at $k' = k_G$ and $k = -k_G$ when light is incident from side 1, while it occurs at $k' = -k_G$ and $k = k_G$ when light is incident from side 2. Note that this behavior is consistent with the requirements stemming from reciprocity: the scattering matrices must be symmetric about the main diagonal, which corresponds to $k=-k'$, or retroreflection (see also **Supplementary Section S5**). Hence, we see that the action of the phase gradient is to shift the response along the retroreflection condition by an amount $(k_G,-k_G)$ in $S_{11}$. Notably, the corresponding shift in $S_{22}$ is $(-k_G,k_G)$, again consistent with reciprocity [symmetry about the dashed line in Fig. 3(h)] and conservation of energy [the dip and peak occurring at the same $k$ in $S_{12}$ and $S_{22}$, and likewise for $S_{11}$ and $S_{21}$]. That is, the behavior of a nonlocal phase gradient is to *tilt* the response in *k*-space [Fig. 3(e)] rather than add a unidirectional momentum kick irrespective of incident direction: it is a vertical asymmetric phenomenon, requiring vertically asymmetry in the structure to achieve (see also the discussion in [48]).

### III.C Nonlocal metalens

While the previous section confirmed that our STCMT extends conventional TCMT modeling to phase gradients, we now demonstrate how this accurate modeling is actually retained even when $k_G$ varies arbitrarily across the device [55]. We consider nonlocal metalenses as a canonical example of spatially varying nonlocal metasurfaces of broad relevance for applications. We first analytically study the eigenwaves of nonlocal metalenses, and then numerically demonstrate excellent agreement with the spatial selectivity demonstrated in Ref. [55]. Our validation with full-wave simulations then motivates us to use STCMT to more thoroughly study nonlocal metalenses with this quick yet accurate tool, shedding more light on their operation. In particular, we demonstrate that, as a function of the band structure $b$, the Q-factor $Q=\omega_0\tau_r$, and the numerical aperture of the metalens $NA$, a nonlocal metalens can transition between two regimes of operation: wavefront-selective and wavefront-shaping. We also explore in **Supplementary Section S10** the eigenwaves for frequencies off the band edge, showing their relation to spatial selectivity and the underlying band structure of the q-BIC.

In this section, we study nonlocal metalenses described by hyperbolic phase gradients of the form

$$\Phi = -k_0 \sqrt{x^2 + f^2}. \tag{48}$$

To find the eigenwave we use relation (33), which uses the terms

$$\rho(x-x',\omega) = \rho_0^*(x-x',\omega) e^{ik_0\sqrt{(x)^2+f^2}} e^{-ik_0\sqrt{(x')^2+f^2}} \\ \rho(x'-x'',\omega) = \rho_0(x'-x'',\omega) e^{ik_0\sqrt{(x')^2+f^2}} e^{-ik_0\sqrt{(x'')^2+f^2}}. \tag{49}$$

The eigenwave of the metalens is therefore the solution to

$$R_{eig} E_{eig}(x,\omega) = \int_{-W/2}^{W/2} dx' \int_{-W/2}^{W/2} dx'' \rho_0^*(x-x',\omega) \rho_0(x'-x'',\omega) e^{ik_0\sqrt{(x)^2+f^2}} e^{-ik_0\sqrt{(x'')^2+f^2}} E_{eig}(x'',\omega). \tag{50}$$

By inspection, as $W \to \infty$ (i.e., ignoring boundary effects), we have

$$E_{eig}(x,\omega) = e^{ik_0\sqrt{x^2+f^2}}, \tag{51}$$

because in this case

$$R_{eig} = \int_{-W/2}^{W/2} dx' \int_{-W/2}^{W/2} dx'' \rho_0^*(x-x',\omega) \rho_0(x'-x'',\omega) = 1, \tag{52}$$

which is identical to the spatially invariant q-BIC modeled by the subscripted kernels. For finite $W$ and radiative boundaries, Eqn. (52) does not yield a perfectly unity result, as expected for a finite device and as discussed in **Supplementary Section S8**. Nevertheless, the idealized solution (51) captures the underlying idea: this nonlocal metalens is selective to a point source set at $z = f$, just as described in Ref. [55]; the eigenwave is the condition in which the phase profile is cancelled everywhere across the device, yielding the same reflectance analogous to a normal incident plane wave exciting a periodic grating.

However, our quantitative definition of an eigenwave introduced here allows us to numerically describe non-ideal eigenwaves for finite structures, including radiative boundary conditions. To do so, we employ again the discrete matrix form of our STCMT. The nonlocal reflection matrix for this scenario is seen in Fig. 4(a,b), from which the eigenwave $E_{eig}$ is computed, depicted in Fig. 6(c) overlaid with the reflected wave ($PE_{eig}$), which is the time-reversed copy of the input as expected.

We also may numerically compute the reflectance due to waves incident from side 1 with

$$E_{in}(x,\omega) = \exp\left(ik_0\sqrt{(x-x_0)^2 + z_0^2}\right). \tag{53}$$

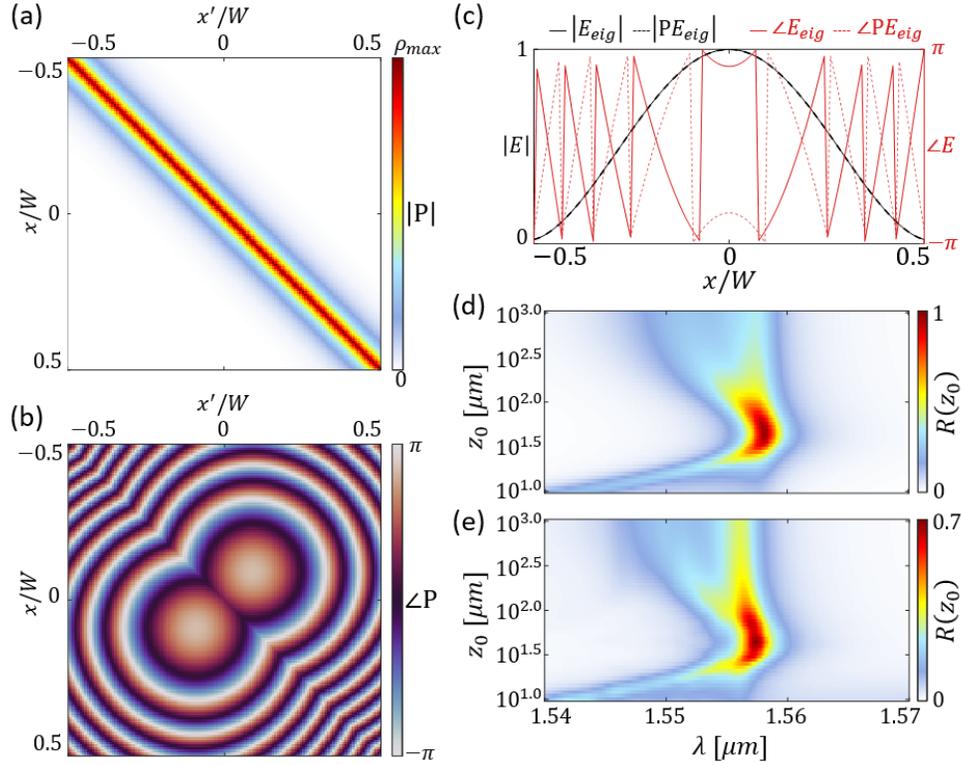

**Figure 4. Nonlocal metalens.** Magnitude (a) and phase (b) of the nonlocal reflection kernel $P$ for a metalens of width $W = 50\mu m$ with q-BIC parameters $b = 0.032\mu m$, $\lambda_0 = 1.558\mu m$, and $\tau_r = 297\mu m$, and $NA = 0.48$, matched to the system in Ref. [55]. (c) Eigenwave and reflected wave, showing unchanged amplitude and conjugated phase profiles upon reflection. (d) The reflectance due to excitation by point sources placed at locations $z_0$ on the optical axis $(x,y) = (0,0)$ for STCMT model. (e) Full-wave simulations reported in Ref. [55].

These waves are idealized point sources placed at locations $(x_0, -z_0)$ relative to the metasurface centered at $(0,0)$. For now, we keep $x_0 = 0$ fixed and vary $z_0$. Using the metasurface introduced in Ref. [55], we numerically compute the reflectance as a function of wavelength and $z_0$ [Fig. 4(d)], and find remarkable agreement with the results from full-wave simulation in Fig. 4(e). The peak reflectance in the STCMT is near-unity, while full-wave simulations predict peak reflectance near 70%. This shows that the metalens has room for optimization in its design. Regardless, our STCMT can clearly capture the response of the nonlocal metalens.

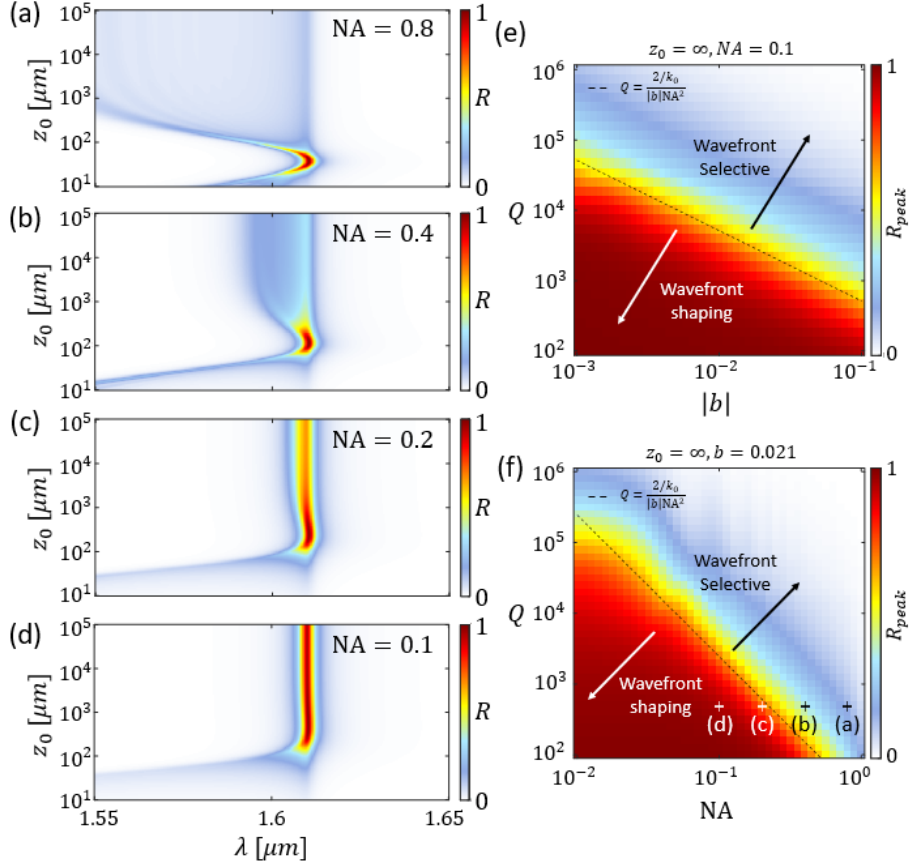

**Figure 5. Selectivity of a nonlocal metalens.** (a-d) Reflectance for a device with $W = 75 \mu m$ and varying NA, with q-BIC parameters $b = 0.021 \mu m$, $\lambda_0 = 1.6104 \mu m$, and $\tau_r = 250 \mu m$, due to excitation by point sources placed at locations $z_0$ on the optical axis $(x, y) = (0, 0)$. (e) Peak reflectance due to a normally incident plane wave for a device with $NA = 0.1$ as a function of $b$ and $Q$. (f) Peak reflectance due to a normally incident plane wave for a device with $b = 0.021$ as a function of NA and $Q$.

Next, we explore the response of the system in Fig. 2 as a function of $b$, $Q$, and numerical aperture,

$$NA = n \frac{W/2}{\sqrt{(W/2)^2 + (f/2)^2}},$$ where $n$ is the index of refraction of the surrounding medium. The NA characterizes the range of deflection angles encoded into the q-BIC, and hence it also characterizes the range of resonant wavelengths across the device for normally incident light. As the NA increases, the part of the nonlocal metalens resonating at a given frequency is reduced, and the reflectance also

diminishes. In other words, as expected, spatial selectivity increases with the lens numerical aperture.

This feature is confirmed numerically in Figs. 5(a-d), which show the spectral reflectance as a function of $z_0$ for four values of NA. When NA=0.8, representing a very large range of deflection angles, only point sources placed near $z_0 = f$ reflect substantial power at the band-edge frequency. Point sources near the eigenwave experience large reflectivity, while point sources away from the eigenwave are transmitted through without resonant interactions with the device. In other words, the nonlocal metalens is strongly spatially selective, and the eigenwave is the optimally selected wavefront to trigger the resonance. When the NA lowers, we observe an increase in reflectance away from the eigenwave condition, e.g., for large $z_0$, which approaches a plane wave in the limit that $z_0 \to \infty$. This study suggests that, when $\text{NA} = 0.8$, we have a wavefront-selective device, only reflecting when the incident light closely matches the eigenwave condition. Yet when $\text{NA} = 0.1$, we have a wavefront-shaping device, abiding large reflectance under a larger range of incident conditions. Notably, away from the eigenwave condition the response to plane waves, $z_0 \gg f$, is better than the response of point sources near the device, $z_0 \ll f$, and it plateaus after a certain distance. This is because point sources near the device launch a large range of spatial frequencies (k-vectors), while point sources far from the device (nearly plane waves) launch a very narrow range. The eigenwave may be considered the condition in which the spatial frequencies launched by the source and encoded the device are matched; the NA is characteristic of the range of k-vectors encoded by the device. Since plane waves represent a lower bound on range of k-vectors, the response plateaus for $z_0 \gg f$ to a value function of the NA.

Importantly, the transition from wavefront-selective to wavefront-shaping is also a function of the degree of locality. When the nonlocality length is substantially larger than the free-space wavelength, the device is highly wavefront-selective, i.e., it only acts as a wavefront-shaping device for very low NA. But when the nonlocality length is comparable to the free-space wavelength, i.e., it supports a large degree of a localization, the device is highly wavefront-shaping, i.e., it acts as a wavefront-shaping device for a large range of NA. We find (**Supplementary Section S11**) that the wavefront-shaping regime is satisfied when

$$\text{NA} < \sqrt{\frac{2/k_0}{|b|Q}} , \quad (54)$$

or equivalently

$$\text{NA} < \frac{1}{\sqrt{2}\pi} \frac{\lambda_0}{\xi_0} . \quad (55)$$

We therefore may quantitatively conclude that the degree of spatial selectivity is due to the degree of nonlocality of the q-BIC. **Supplementary Section S12** further quantifies this feature by discussing the expansion in the eigenbasis of the device, an analysis related to modal expansions in optics [64].

We now explore devices with varying $b$, $Q$, and NA to demonstrate the transition between wavefront-selective and wavefront-shaping recipes. Figure 5(e) shows the peak reflectance for a normally incident plane wave $z_0 = \infty$ for a device with NA $= 0.1$ as a function of $b$ and $Q$: when they are both large, i.e., a highly nonlocal device, the plane wave is transmitted unperturbed through the device, indicating wavefront-selectivity. When they are both small, i.e., a highly local device, the plane wave is anomalously reflected with near-unity efficiency, and the device is wavefront-shaping. The transition Q-factor for a given value of $b$ is overlaid on Fig. 5(e) by rearranging Eqn. (54), showing good agreement between the numerical calculations and the physical arguments outlined here. Additionally, Fig. 5(f) shows a device with $b = 0.021 \mu m$ as a function of NA and $Q$, showing similar agreement in the transition between the two regimes. The individual scenarios in Fig. 5(a-d) are marked in Fig. 5(f).

## IV. Thermal metasurfaces

As an additional relevant application, we use STCMT to model thermal metasurfaces [56],[66]. Engineered nonlocalities have been recently shown to provide a powerful platform to shape thermal emission and photoluminescence from metasurfaces, and STCMT may enable rapid study and prototyping of focused incoherent light emission with respect to the defined nonlocality length and the design parameters. To tackle this problem, we need to extend the STCMT formulation in three ways: we (i) account for polarization ports in a reflection- and absorption-only scattering problem, (ii) incorporate a spatially varying local metasurface response in the background scattering matrix $C(x)$, and (iii) add absorption loss, distinguishing the radiative scattering rate $\gamma_r$ and the nonradiative scattering (loss) rate $\gamma_{nr}$. Our scattering matrix $S_{ij}$ is now indexed by linear polarizations $i, j \in x, y$, and we allow the background scattering to follow an idealized geometric phase metasurface implemented by pillars with orientation angles $\theta(x)$:

$$C(x) = \begin{bmatrix} \cos[2\theta(x)] & \sin[2\theta(x)] \\ \sin[2\theta(x)] & -\cos[2\theta(x)] \end{bmatrix}. \tag{56}$$

Meanwhile, for perturbations governing the q-BIC defined by an orientation angle $\alpha$, the coupling coefficients are [66]

$$|d(x)\rangle = \sqrt{\gamma_r} \left\{ \begin{bmatrix} \cos[2\alpha(x)] \\ \sin[2\alpha(x)] \end{bmatrix} - i \begin{bmatrix} \cos[2\theta(x) - 2\alpha(x)] \\ \sin[2\theta(x) - 2\alpha(x)] \end{bmatrix} \right\}. \tag{57}$$

Finally, to account for the presence of loss, we make the simple adjustment

$$\xi(\omega) = \sqrt{\frac{ib/2}{\gamma_r + \gamma_{nr} + i(\omega_0 - \omega)}}. \tag{58}$$

For clarity, we may also separate the scattering kernel into its local and nonlocal components

$$\sigma(x, x', \omega) = \rho_{loc}(x) + \rho_{nonloc}(x, x', \omega), \tag{59}$$

where

$$\rho_{loc}(x) = C(x)\delta(x - x') \tag{60}$$

$$\rho_{nonloc}(x, x', \omega) = G(x, x', \omega)|d(x)\rangle\langle d^*(x')|. \tag{61}$$

Then, assuming that the power not reflected is absorbed by the device, the absorption of a given input wave can be simply computed as

$$\mathcal{A}(\omega) = 1 - \langle s_+ | \sigma^*(x,x',\omega)\sigma(x,x',\omega) | s_+ \rangle, \tag{62}$$

while the corresponding thermal emission is described by the time-reversed wave, following the universal modal radiation laws [67].

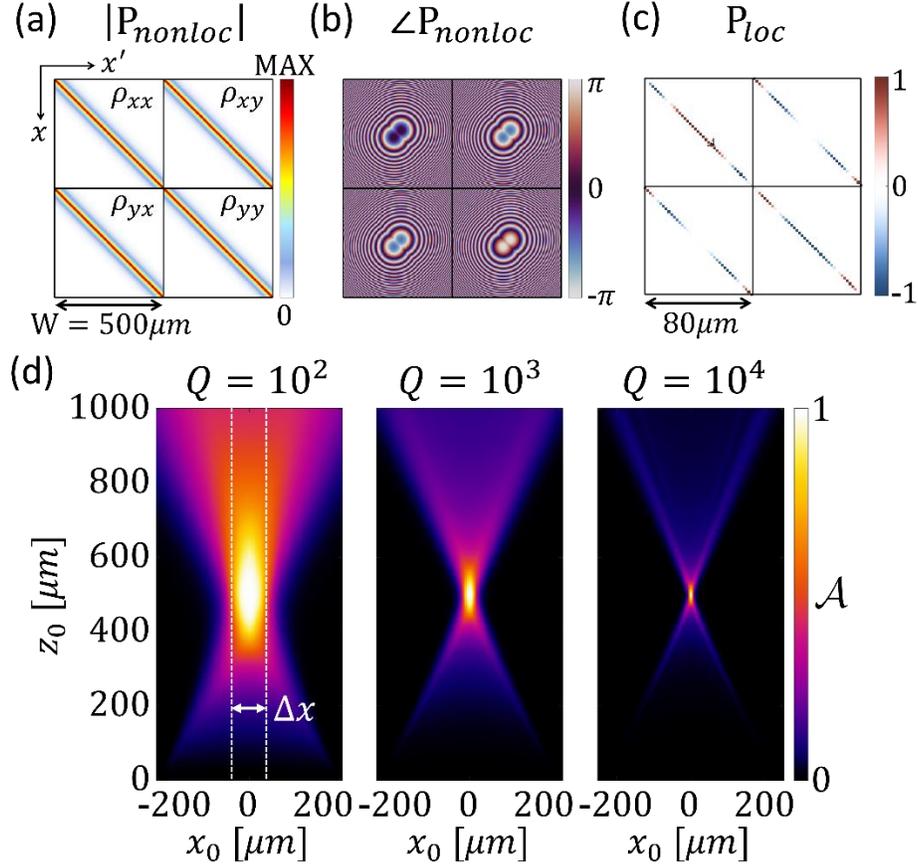

**Figure 6. Thermal metalenses.** (a-b) Magnitude and phase of the nonlocal component of the nonlocal reflection kernel $P_{nonloc}$ at the band edge frequency of a thermal metalens that is $W = 500 \mu m$ wide, focusing to a point $f = 500 \mu m$ away with $\lambda_0 = 4 \mu m$. (c) The corresponding local reflection kernel $P_{loc}$ (central $80 \mu m$ depicted for visual clarity). (d) Computed absorption for sample point sources placed at positions $(x_0, z_0)$ due to a thermal lens with $b = 0.15 \mu m$ for three example Q-factors.

As a canonical example, Fig. 6 studies a thermal metalens that preferentially emits to a focal point $f$ with a circular polarization, encoded using the profiles [56]

$$\theta(x) = \pm \frac{2\pi}{\lambda_0}\sqrt{x^2 + f^2} \tag{63}$$

$$2\alpha(x) = \theta(x) \pm \frac{\pi}{4} \tag{64}$$

for the chosen polarization $[1 \quad \mp i]/\sqrt{2}$. Figures 6(a,b) show $\mathrm{P}_{nonloc}$, the matrix form of $\rho_{nonloc}$ for an example thermal metalens, where the four quadrants correspond to the input and output polarization states in an $x, y$ basis. Figure 6(c) shows $\mathrm{P}_{loc}$, the matrix form of $\rho_{loc}$, which in contrast to the lossless, scalar nonlocal metalens of the previous section, encodes a varying birefringent axis along the main diagonals of each quadrant [Eqn. (56) with Eqn. (63)]. Then, we compute the absorption of the metalens due to excitation by point sources placed at positions $(x_0, z_0)$ above the metasurface. Figure 6(d) shows the spatial absorption maps for three example Q-factors operating at the band-edge frequency, showing that as the lifetime of the q-BIC grows, the focal spot tightens. This directly corresponds to an increase in the nonlocality length $\xi_0 = \sqrt{b\tau_r}$, i.e., the metasurface becomes increasingly wavefront selective; increasing $b$ has the same effect at the band-edge frequency. **Supplementary Section S13** discusses the response as a function of incident frequency.

Beyond specific examples, we may rapidly compute the full width at half maximum (FWHM), $\Delta x$ of the thermal metalens response as a function of Q-factor and $NA$. Figure 7(a) shows how $\xi_0$ and the peak absorption $\mathcal{A}_{peak}$ depend on $Q$. Notably, we see a drop off in $\mathcal{A}_{peak}$ around the point where $\xi_0 = W$, i.e., when the nonlocality length is comparable to the aperture size (width) of the lens. This naturally stems from incomplete interference: the finite size of the lens is too small, and the q-BIC leaks out the sides of the metalens instead of building up a complete Fano resonance through destructive interference. Figure 7(b) shows the absorption at the focal plane $z_0 = f$ as a function of $Q$, showing a continuous trend consistent with Fig. 6(d) when $\xi_0 < W$. However, when $\xi_0 > W$ we notice instead that the FWHM plateaus, as confirmed in Fig. 7(c) above a certain $Q$. Below that threshold Q-factor, the target $NA$, together with $Q$, determine $\Delta x$; above that threshold Q-factor, $\Delta x$ is determined only by the $NA$. From this, we learn that increasing the Q-factor does not linearly improve the performance of a thermal metalens. Instead, there is a threshold $Q$ above which the focal spot does not continue to tighten; instead, increasing the Q-factor only reduces the intensity of the response. While the transition near $\xi_0 = W$ is gradual, yielding a 'gray area' rather than a sharp threshold, the distinction between the two regimes is clear. Figure 7 suggests, intuitively, that to improve the response further (i.e., reduce $\Delta x$ while maintaining $\mathcal{A}_{peak} \approx 1$) we must increase the metalens aperture ($W$) while also increasing the coherence across the aperture ($\xi_0$).

## V. Discussion and Conclusion

TCMT is the foundational model and framework for studying infinite, periodic resonant flat optics. In analogy, here we have introduced STCMT as the framework to model space-varying resonant flat optics. Given the success of TCMT, we expect STCMT to become a foundational tool for design nonlocal meta-optics before implementing a concrete geometry in full-wave simulations and experiments. STCMT qualitatively and quantitatively clarifies the classification of nonlocal metasurfaces into two functionalities: wavefront-selective nonlocal metasurfaces and wavefront-shaping nonlocal metasurfaces. In practice, this allows rapid classification and prototyping of a system for comparison to the selectivity requirements of an application. More generally, STCMT enables computation of the idealized performance of a given system, taking into account finite size, space-varying properties, and degrees of freedom manipulatable across the surface. In other words, it sets goalposts (upper bounds on performance) to be reached for in practical implementation under a given set of assumptions (geometry, number of layers, loss, size, reciprocity, etc.).

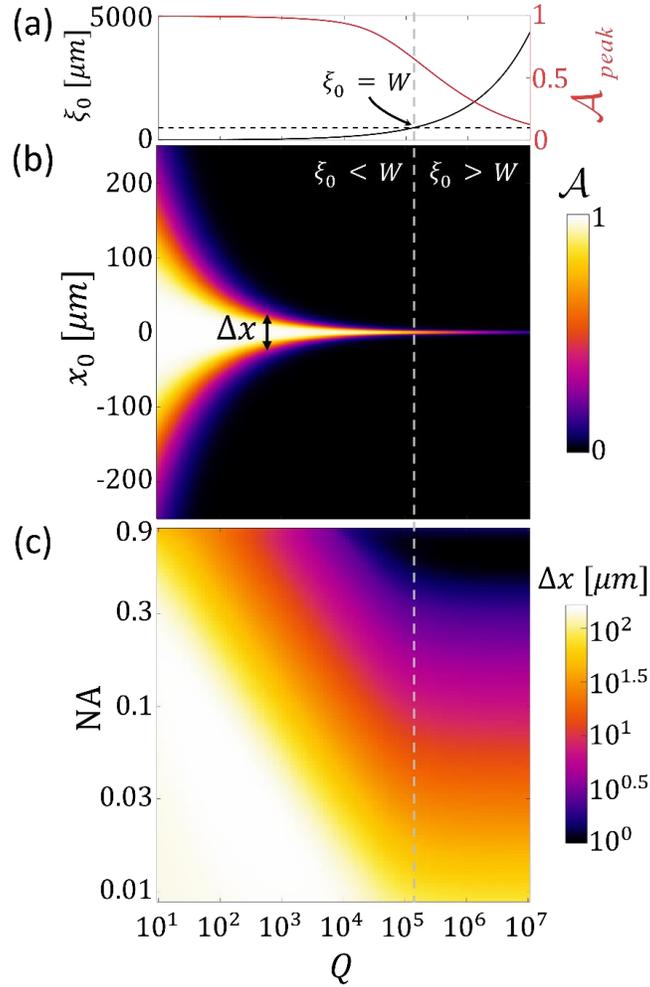

**Figure 7. Effect of nonlocality on thermal metalenses.** (a) Nonlocality $\xi_0 = \sqrt{b\tau_r}$ and peak absorption $\mathcal{A}_{peak}$ [i.e., for a source placed at $(x_0, z_0) = (0, f)$] as a function of $Q$. The horizontal black dashed line marks a length of $W$; the vertical gray dashed line marks the $Q$ at which $\xi_0 = W$. (b) Absorption for point sources placed in the focal plane at $z_0 = f$ as a function of $Q$. (c) Full width at half maximum $\Delta x$ of the response at the focal plane as a function of $Q$ and numerical aperture $NA$, showing two regimes of behavior.

In addition to its practical advantages, among the key benefits of CMT is the potential to add quantitative and qualitative understanding to the underlying physics of a resonant response. STCMT captures and provides insight into several novel aspect of recent efforts on nonlocal metasurfaces: directly stemming from STCMT, we clarified that: (i) the two hallmarks of a nonlocal phase gradient (shift of band structure in input angle, and anomalous reflection) are simple consequences of a mixed

Fourier transformation in the context of STCMT; (ii) a symmetric response after such mixed Fourier transformation is enforced by reciprocity, revealing that a near-unity efficiency nonlocal phase gradient is a vertically asymmetric phenomenon, in turn revealing the requirement of vertical asymmetry in the structure; and (iii) that the spatial selectivity is intrinsically tied to the eigenwave encoded into the metasurface, to a degree controlled by the nonlocality length $\xi_0$ of the q-BIC (a quantitative measure of the lateral interaction length, thereby setting the minimum metasurface size required to observe a Fano response, as seen in Fig. 7(c)).

While we demonstrated the utility of STCMT for understanding and studying diffractive nonlocal metasurfaces, we emphasize that this represents a unified framework capable of describing metasurfaces *with or without* the assumption of locality. For instance, the local assumption is recovered if only the main diagonal of the nonlocal matrices is populated, and the effects of nearest neighbors are included when more elements than just the main diagonal are nonzero. As discussed further in **Supplementary Section S14**, our framework may enable future work explicitly taking into account neighbor-neighbor (nonlocal) interactions—that is, to go beyond the so-called "local phase approximation" foundational to metasurface design (perhaps by extension to quasi-normal modes [65]). Additionally, our framework is readily adapted to include both local and nonlocal metasurface behaviors simultaneously, allowing rapid study and prototyping of thermal metasurfaces originally reported in Ref. [56] and recently experimentally demonstrated in Ref. [66]. Future directions may also extend our results to study the resonant dynamics of several spectrally overlapping nonlocal modes (as in Ref. [7]), leaky-wave metasurfaces based on travelling wave q-BICs [67], the effects due to partial coherence (i.e., study the effects of nonlocal scattering on the cross-spectral density [69]), and even nonreciprocal behavior [70]. Finally, **Supplementary Section S15** discusses the limitations of the present study.

In summary, in this work we have extended the coupled mode theory formalism to capture the resonant physics of space-varying metasurfaces. We demonstrated excellent agreement with full-wave simulations of devices within the rapidly emerging paradigm of diffractive nonlocal metasurfaces, which are capable of wavefront and thermal emission manipulation via the physics of q-BICs. This emerging platform is capable of unprecedent control of light for applications such as augmented reality, secure optical communications, compact optical sources, and nonlinear and quantum optics. STCMT elegantly captures the essential physics of these systems, and enables rapid, analytical and numerical study and design of next-generational meta-optical system.

### Acknowledgements

This work was supported by AFOSR (FA9550-18-1-0379) and the Simons Foundation. We thank Nanfang Yu and Stephanie Malek for discussions on the regime of wavefront-shaping metasurfaces.

# Supplementary Materials for:

# Spatio-temporal coupled mode theory for nonlocal metasurfaces

Adam Overvig[1], Sander Mann[1], and Andrea Alù[1,2,*]

[1]Photonics Initiative, Advanced Science Research Center, City University of New York, New York, NY 10031, USA
[2]Physics Program, Graduate Center of the City University of New York, New York, NY 10016, USA
*Corresponding author: aalu@gc.cuny.edu

## S1. Temporal coupled mode theory (TCMT)

### S1a. General development of TCMT

Periodic nonlocal metasurfaces are often studied for their resonant properties, which are well-captured by temporal coupled mode theory (TCMT). Here, we review the fundamentals of TCMT and compare the numerical model with full-wave simulations of a conventional, periodic q-BIC device. In this section, the TCMT begins by equations of motion in the time-domain, assumes time-harmonic solutions to yield equations in the frequency domain, and then Taylor expands the resonant frequency as a function of momentum to yield scattering coefficients in the momentum-frequency domain. The main results of the paper are achieved by extending this approach to the space-frequency domain using suitable Fourier transformation and spatial adjustment of the q-BIC coupling properties.

We begin with the equation of motion for a q-BIC with complex modal amplitude $a(t)$ excited by an incoming wave $|s_+\rangle$, where $|a(t)|^2$ is normalized to be the energy stored in the q-BIC per unit area at time $t$ and $\langle s_+|s_+\rangle$ the incident intensity:

$$\frac{da(t)}{dt} = -(i\omega_r + \gamma_r + \gamma_i)a(t) + \langle \kappa^* | s_+ \rangle. \tag{1}$$

The coupling vector $|\kappa\rangle$ contains the coupling coefficients from free-space excitation, $\omega_r$ is the (resonant) modal frequency of the q-BIC, $\gamma_r$ is the radiative scattering rate, and $\gamma_i$ is the nonradiative scattering rate (which we will take to be 0 throughout). We note that the scattering rates may be used interchangably with a corresponding lifetime, e.g. $\tau_r = 1/\gamma_r$ is the radiative lifetime of the mode. We also wish to know the outgoing (scattered) waves $|s_-\rangle$, which are also normalized so that $\langle s_- | s_- \rangle$ is the outgoing intensity. We assume time-harmonic solutions of the form

$$\begin{aligned} a(t) &= a_0 \exp(-i\omega t) \\ |s_+\rangle &= \exp(-i\omega t)|s_{0+}\rangle, \\ |s_-\rangle &= \exp(-i\omega t)|s_{0-}\rangle \end{aligned} \tag{2}$$

yielding

$$\langle \kappa^* | s_{0+} \rangle = [i(\omega - \omega_r) - \gamma_r]a_0 \tag{3}$$

Then, if in the absence of the q-BIC the outgoing wave is given by

$$|s_{0-}\rangle = C(\omega)|s_{0+}\rangle, \tag{4}$$

where $C(\omega)$ is the background scattering matrix, the contribution from the decaying q-BIC means $|s_-\rangle$ must be given by

$$|s_{0-}\rangle = C(\omega)|s_{0+}\rangle + |d\rangle a_0. \tag{5}$$

Equations (3) and (5) form the general basis of the TCMT in the frequency domain [1],[6]. A more specific TCMT is then developed by (i) applying relevant physical constraints and (ii) parameterizing the system of interest with phenomenological parameters. In doing so, we will retrieve the scattering matrix $S$ relating the incoming waves to the the outgoing waves,

$$\left|s_{0-}\right\rangle = S\left|s_{0+}\right\rangle, \tag{6}$$

parameterized by quantities rationally controlled through the symmetry properties of q-BICs. Regarding (i), here we are interested in physical systems subject to the constraints of reciprocity, energy conservation, and time-reversal [6]. Reciprocity demands that the coupling into the q-BIC must be equivalent to coupling out, namely,

$$\left|\kappa\right\rangle = \left|d\right\rangle. \tag{7}$$

Energy conservation means that the intensity coupling out of the q-BIC is constrained by the radiative decay rate, in particular,

$$\left\langle d \middle| d \right\rangle = 2\gamma_r. \tag{8}$$

Finally, for time reversal invariant systems, the equations describing a decaying q-BIC with no driving (incoming) waves must, under time reversal, match the equations describing a q-BIC being populated at the radiative decay rate without scattering to any outgoing waves, which requires

$$C\left|d^*\right\rangle = -\left|d\right\rangle. \tag{9}$$

We may now eliminate the modal amplitude in Eqns. (3) and (5), and then apply the reciprocity condition in Eqn. (7) to obtain the scattering matrix

$$S = C + \frac{|d\rangle\langle d^*|}{i(\omega - \omega_r) - \gamma_r}. \tag{10}$$

## S2a. TCMT for the device class of interest

We are interested in particular in the local scattering matrix of the form:

$$C = e^{i\Phi_c} \begin{bmatrix} r_0 & 0 & -it_0 & 0 \\ 0 & r_0 & 0 & -it_0 \\ -it_0 & 0 & r_0 & 0 \\ 0 & -it_0 & 0 & r_0 \end{bmatrix}, \tag{11}$$

where $r_0$ and $t_0$ are real-valued Fresnel coefficients (e.g., as approximated by a thin film interference model). This scattering matrix uses a basis for the fields of the form:

$$E = \begin{bmatrix} E_{1,x} \\ E_{1,y} \\ E_{2,x} \\ E_{2,y} \end{bmatrix}, \tag{12}$$

where $E_{i,p}$ is the electric field on side $i$ with polarization $p$. Then we seek the complex coupling coefficients

$$|d\rangle = \begin{bmatrix} d_1 \\ d_2 \\ d_3 \\ d_4 \end{bmatrix} e^{i\Phi_c/2} \tag{13}$$

where we include the background scattering phase factor for convenience, allowing us to write Eqn. (10) as

$$S = e^{i\Phi_c} \left\{ \begin{bmatrix} r_0 & 0 & -it_0 & 0 \\ 0 & r_0 & 0 & -it_0 \\ -it_0 & 0 & r_0 & 0 \\ 0 & -it_0 & 0 & r_0 \end{bmatrix} + \frac{1}{i(\omega-\omega_0)-\gamma_r} \begin{bmatrix} d_1 d_1 & d_1 d_2 & d_1 d_3 & d_1 d_4 \\ d_2 d_1 & d_2 d_2 & d_2 d_3 & d_2 d_4 \\ d_3 d_1 & d_3 d_2 & d_3 d_3 & d_3 d_4 \\ d_4 d_1 & d_4 d_2 & d_4 d_3 & d_4 d_4 \end{bmatrix} \right\} \quad (14)$$

The development of the TCMT now proceeds by (1) writing down a phenomenological form of $|d\rangle$ relevant to the geometry in Fig. S1 and then (2) applying the constraints of conservation of energy, coordinate system invariance, and time-reversal symmetry to determine the complex amplitudes in $|d\rangle$. We begin with step (1) guided by the selection rules. Seen in Fig. S1(a), we study the q-BIC system from Refs. [3]-[55], wherein a symmetric photonic crystal slab, sandwiched between an identical superstrate and substrate, is perturbed distinctly at the top and bottom interfaces. In the absence of this perturbation, the system is two-port symmetric and the q-BIC is fully bound. The presence of a perturbation breaks this symmetry, meaning that scattering to free-space is introduced in such a way that the scattering to sides $p=1$ and $p=2$ (bottom and top, respectively) distinct, in principle. This perturbation is characterized by four parameters: the magnitude of the perturbation at the bottom and top interfaces are $\delta_1$, and $\delta_2$, and specify how deviated the ellipses are from a perfect circle; the orientation angles of the ellipses are characterizes by angles $\alpha_1$ and $\alpha_2$ for the bottom and top interfaces while the second ellipse in each layer is oriented at an angle $\alpha_1 + 90°$ and $\alpha_2 + 90°$, respectively. As shown in Fig. S1(d), we describe this system as follows: the perturbation at the bottom interface scatters light to polarization $\phi$ with a strength proportional to the magnitude of the perturbation, $\delta_1$; the perturbation at the top interface scatters light to polarization $\theta$ with a strength proportional to the magnitude of that perturbation, $\delta_2$. As shown in Ref. [3],[4], the polarization angles follow closely the relations

$$\begin{aligned}\phi &\approx 2\alpha_1\\ \theta &\approx 2\alpha_2\end{aligned}, \tag{15}$$

and, without loss of generality, we may parameterize the perturbation strengths as

$$\begin{aligned}\delta_1 &= \delta_0 \cos(\delta)\\ \delta_2 &= \delta_0 \sin(\delta)\end{aligned} \tag{16}$$

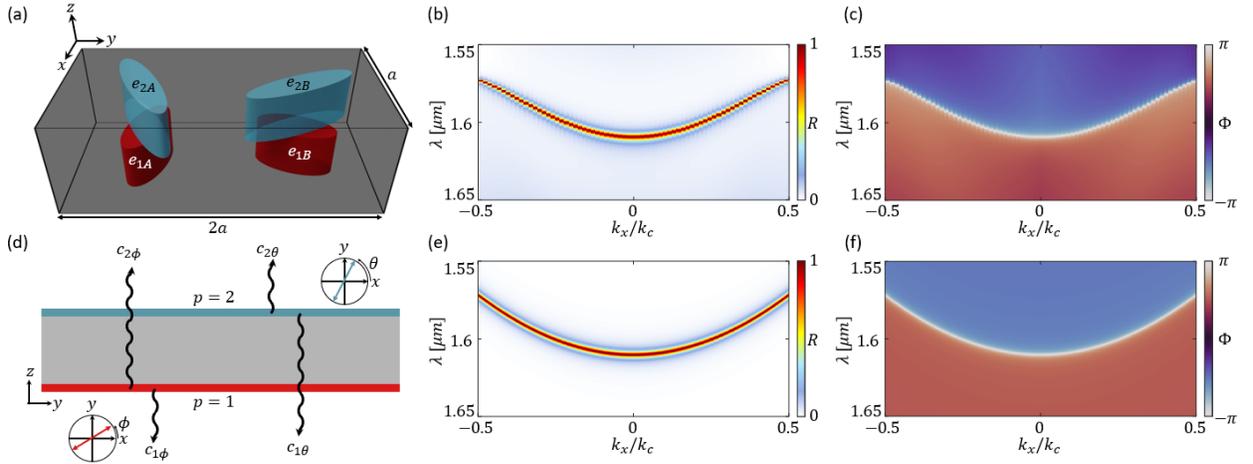

**Figure S1.** Comparison of a nonlocal metasurface implementation studied by full-wave simulations (a-c) and modeled by TCMT (d-f). (a) Unit cell of a nonlocal metasurface constructed from four elliptical inclusions. (b,c) Reflection and reflected phase for the case in (a) where the ellipses are aligned in the. (d) Schematic of the TCMT model of the device in (a). (e,f) Reflection and reflected phase from a TCMT fit to match the device in (a-c) near $k_x = 0$.

where $\delta_0$ is a measure of the overall perturbation strength and $\delta$ is an angular parameter determining the relative contribution from each interface. By construction (or definition for a q-BIC), the Q-factor (or radiative lifetime) is controlled by the overall perturbation strength, and hence we expect $\tau_r \propto 1/\delta_0^2$. And as shown in Ref. [4], $\delta$ plays an important role in adjusting the overall phase of $|d\rangle$, motivating this parameterization.

With these assumptions, we may write the components of $|d\rangle$ based on the sum of the scattering events from each interface:

$$\begin{aligned} d_1 &= c_{1\phi}\cos(\phi) + c_{1\theta}\cos(\theta) \\ d_2 &= c_{1\phi}\sin(\phi) + c_{1\theta}\sin(\theta) \\ d_3 &= c_{2\phi}\cos(\phi) + c_{2\theta}\cos(\theta) \\ d_4 &= c_{2\phi}\sin(\phi) + c_{2\theta}\sin(\theta) \end{aligned} \quad (17)$$

where the coefficients $c_{ip}$ represent the coupling to port $i$ from the scatterer producing polarization $p$. These coefficients are not completely independent, per the symmetry of the device. In particular, we may write them in the form

$$\begin{aligned} c_{1\phi} &= d_0 \cos(\delta) e^{i\Phi} \\ c_{2\theta} &= d_0 \sin(\delta) e^{i\Phi} \\ c_{2\phi} &= c_{1\phi} g e^{i\Delta\Phi} \\ c_{1\theta} &= c_{2\theta} g e^{i\Delta\Phi} \end{aligned}, \quad (18)$$

where the first two coefficients represent scattering directly from an interface to their respective ports and the latter two represent indirect scattering from an interface through the slab and to the opposite port. The direct scattering coefficients are not equal in magnitude because the perturbation strengths are different (parameterized by $\delta$), but by the symmetry of the unperturbed system (and thereby, the mode profile), a symmetric mode will scatter with the same direct phase $\Phi$. An anti-symmetric mode will scatter with opposite phase, which may accounted for with $\delta$. The indirect scattering coefficients [the latter two in Eqn. (18)] may be written without loss of generality in terms of their direct counterparts and a relative amplitude $g$ and phase factor $\Delta\Phi$. In our case, the phase factor is $\Delta\Phi = -\pi/2$, following the phase relation of reflected (direct) and transmitted (indirect) light in Eqn. (11) for the vertically symmetry (unperturbed) structure.

We now begin step (2) of developing the TCMT by applying physical constraints by applying the consequence of energy conservation,

$$\langle d | d \rangle = 2\gamma_r \tag{19}$$

and time-reversal symmetry,

$$C | d^* \rangle = -| d \rangle, \tag{20}$$

to determine the allowed forms the complex coefficients. From Eqn. (19) we have

$$|d_1|^2 + |d_2|^2 + |d_3|^2 + |d_4|^2 = 2\gamma_r \tag{21}$$

which, upon inserting Eqn. (17) and Eqn. (18), leads to

$$|d_0|^2 = \frac{2\gamma_r}{1+g^2}. \tag{22}$$

Meanwhile, from time reversal symmetry [Eqn. (20)], we have the constraint

$$r_0 d_1^* - it_0 d_3^* = -d_1. \tag{23}$$

from which we have

$$\frac{r_0\left(\cos(\phi)\cos(\delta)+i\cos(\theta)\sin(\delta)g\right)-it_0\left(i\cos(\phi)\cos(\delta)g+\cos(\theta)\sin(\delta)\right)}{\cos(\phi)\cos(\delta)-i\cos(\theta)\sin(\delta)g}=-e^{2i\Phi}. \tag{24}$$

Calling the left-hand side $h_0$, the right-hand side requires that $|h_0|=1$, from which it may be shown that

$$g=\frac{1-|r_0|}{\sqrt{1-r_0^2}}. \tag{25}$$

To simplify the final form, we now note that for the system at study here, we operate such that the resonant frequency $\omega_r$ is near the transmission peak of a Fabry-Perot background, $r_0=0$, yielding

$$\begin{aligned} g &= 1 \\ h_0 &= 1 \\ e^{i\Phi} &= i \\ |d_0| &= \sqrt{\gamma_r} \end{aligned} \tag{26}$$

And thereby, we arrive at the following form for the scattering coefficients,

$$\begin{aligned} d_1 &= i\sqrt{\gamma_r}\left[\cos(\phi)\cos(\delta)-i\cos(\theta)\sin(\delta)\right] \\ d_2 &= i\sqrt{\gamma_r}\left[\sin(\phi)\cos(\delta)-i\sin(\theta)\sin(\delta)\right] \\ d_3 &= i\sqrt{\gamma_r}\left[-i\cos(\phi)\cos(\delta)+\cos(\theta)\sin(\delta)\right] \\ d_4 &= i\sqrt{\gamma_r}\left[-i\sin(\phi)\cos(\delta)+\sin(\theta)\sin(\delta)\right] \end{aligned} \tag{27}$$

Equation (27) may be written in a more suggestive form by separating the upwards and downward scattering coefficients,

$$d_{up} = \begin{bmatrix} d_1 \\ d_2 \end{bmatrix} = i\sqrt{\gamma_r}\left\{\cos(\delta)\begin{bmatrix} \cos(\phi) \\ \sin(\phi) \end{bmatrix} - i\sin(\delta)\begin{bmatrix} \cos(\theta) \\ \sin(\theta) \end{bmatrix}\right\}$$
$$d_{down} = \begin{bmatrix} d_3 \\ d_4 \end{bmatrix} = i\sqrt{\gamma_r}\left\{-i\cos(\delta)\begin{bmatrix} \cos(\phi) \\ \sin(\phi) \end{bmatrix} + \sin(\delta)\begin{bmatrix} \cos(\theta) \\ \sin(\theta) \end{bmatrix}\right\}, \quad (28)$$

That is, relative to the factor $i$, real part of the upward scattered state's Jones vector is given by the linear polarization state $\phi$ with strength $\cos(\delta)$ while the imaginary part is given by the linear polarization state $\theta$ with strength $g\sin(\delta)$.

With Eqn. (27) in hand, the elements of $|d\rangle$ are known, and the overall scattering matrix is uniquely determined by the phenomenological parameters $r, \omega_0, \tau_r, \phi, \theta$. We may write the final scattering matrix in the form:

$$S = e^{i\Phi_c}\left[C + \frac{1}{1 - i\Omega\tau_r}D\right], \quad (29)$$

with our case specifically parameterized by

$$C = -i \begin{bmatrix} 0 & 0 & 1 & 0 \\ 0 & 0 & 0 & 1 \\ 1 & 0 & 0 & 0 \\ 0 & 1 & 0 & 0 \end{bmatrix},$$

$$D = \begin{bmatrix} d_1 \\ d_2 \\ d_3 \\ d_4 \end{bmatrix} \begin{bmatrix} d_1 & d_2 & d_3 & d_4 \end{bmatrix},$$

$$\begin{bmatrix} d_1 \\ d_2 \\ d_3 \\ d_4 \end{bmatrix} = \cos(\delta) \begin{bmatrix} \cos(\phi) \\ \sin(\phi) \\ -i\cos(\phi) \\ -i\sin(\phi) \end{bmatrix} + \sin(\delta) \begin{bmatrix} -i\cos(\theta) \\ -i\sin(\theta) \\ \cos(\theta) \\ \sin(\theta) \end{bmatrix}, \quad (30)$$

$$\Omega = \omega - \omega_r,$$

$$\Phi_c = \pi,$$

where we note that the frequency difference $\Omega$ may be adjusted to account for nonradiative loss rate $\gamma_i$ by the replacement

$$\Omega \to \Omega = \omega - (\omega_0 - i\gamma_i). \quad (31)$$

These phenomenological parameters are each directly controlled by geometric (and material) parameters as

$$\begin{aligned} \omega_0 &= f(\epsilon) \\ \tau_r &\propto 1/\delta_0^2 \\ \phi &\approx 2\alpha_1 \\ \theta &\approx 2\alpha_2 \end{aligned} \quad (32)$$

where $f(\epsilon)$ represents some function (with no generalized closed form) of the period, permittivity, and other geometrical parameters of the unperturbed PCS.

Finally, we may simply extend this model to the momentum-frequency domain by the replacement $\omega_r \rightarrow \omega_r(\mathbf{k})$ and by taking the incoming and outgoing wave basis as planewaves with in-plane momenta $\mathbf{k}$. Here, near normal incidence (i.e., a band edge) the resonant frequency may be approximated parabolically as

$$\omega_r(k) \approx \omega_0 + \frac{b}{2}k^2, \qquad (33)$$

where $\omega_0$ is the resonant frequency at normal incidence (i.e., the band-edge frequency) and *b* is the Taylor expansion coefficient describing the curvature of the band near the band edge. Figure S1(b) shows the reflection calculated by the FDTD, and Figure S1(e) shows the reflection using the TCMT model fit to the response of this device. Similarly, the phase upon reflection is shown in Fig. S1(c) for FDTD and Fig. S1(f) for TCMT. This excellent agreement validates the use of the analytical form, and motivates its extension to cover the more interesting spatially varying cases shown in Ref. [2].

**S1c. Circular polarization**

In the main text, we are principally interested in the case where $\delta = \pi/4$ and $\theta = \phi + \pi/2$. In this case, Eqn. (28) becomes

$$d_{up} = \begin{bmatrix} d_1 \\ d_2 \end{bmatrix} = i\sqrt{\gamma_r/2} \left\{ \begin{bmatrix} \cos(\phi) \\ \sin(\phi) \end{bmatrix} - i\begin{bmatrix} \cos(\phi+\pi/2) \\ \sin(\phi+\pi/2) \end{bmatrix} \right\}$$
$$d_{down} = \begin{bmatrix} d_3 \\ d_4 \end{bmatrix} = i\sqrt{\gamma_r/2} \left\{ -i\begin{bmatrix} \cos(\phi) \\ \sin(\phi) \end{bmatrix} + \begin{bmatrix} \cos(\phi+\pi/2) \\ \sin(\phi+\pi/2) \end{bmatrix} \right\} \qquad (34)$$

which may be written as

$$d_{up} = \begin{bmatrix} d_1 \\ d_2 \end{bmatrix} = i\sqrt{\gamma_r/2} \begin{bmatrix} 1 \\ -i \end{bmatrix} \exp(i\phi)$$
$$d_{down} = \begin{bmatrix} d_3 \\ d_4 \end{bmatrix} = i\sqrt{\gamma_r/2} \begin{bmatrix} -i \\ 1 \end{bmatrix} \exp(-i\phi)$$

(35)

In a circular polarization basis, we therefore have

$$d_{up} = i\sqrt{\gamma_r/2} \begin{bmatrix} 1 \\ 0 \end{bmatrix} \exp(i\phi)$$
$$d_{down} = i\sqrt{\gamma_r/2} \begin{bmatrix} -i \\ 0 \end{bmatrix} \exp(-i\phi)$$

(36)

as used in the main text.

## S2. Obtaining the STCMT dynamical equations

### S2a. Using a Taylor Expansion

Well within the purview of TCMT is modelling infinitely periodic structure defined by a band structure in momentum-frequency space. In this case, simple transformation yields:

$$i[\Omega - \omega]a(\omega) = \langle \kappa^* | s_+(\omega) \rangle \tag{37}$$

$$|s_-(\omega)\rangle = C(\omega)|s_+(\omega)\rangle + a(\omega)|d\rangle \tag{38}$$

where we introduce the complex resonant frequency $\Omega = \omega_0 + i\gamma$, and assume that the time dependence of the broadband, background scattering is instantaneous, $C \propto \delta(t-t')$ (see the discussion in Ref. [6]). Next, we explicitly include the momentum dependence

$$i[\Omega(k) - \omega]a(k,\omega) = \langle \kappa^*(k) | s_+(k,\omega) \rangle \tag{39}$$

$$|s_-(k,\omega)\rangle = C(k,\omega)|s_+(k,\omega)\rangle + a(k,\omega)|d(k)\rangle. \tag{40}$$

To proceed, we assume the Taylor expanded form of the complex resonant frequency:

$$\Omega(k) = \left(\omega_0 + ck + \frac{b}{2}k^2 + \ldots\right) + i\left(\gamma_0 + \frac{\gamma_1}{2}k^2 + \ldots\right) \quad (41)$$

With Eqn. (41), Fourier transforming Eqns. (39) and (40) gives

$$\frac{da(x,t)}{dt} + i(\omega_0 + i\gamma)a(x,t) - c\frac{da(x,t)}{dx} + i(b + i\gamma_1)\frac{d^2 a(x,t)}{dx^2} = \int dx' \langle \kappa(x,x') | s_+(x') \rangle \quad (42)$$

$$|s_-(x)\rangle = \int dx' \left( C(x,x') | s_+(x') \rangle + a(x) | d(x,x') \rangle \right) \quad (43)$$

On the left-hand side, the spatial derivative terms are obtained from the well-known Fourier transform properties that the nth derivative satisfies

$$\mathbf{F}\left\{\frac{d^n}{dx^n} f(x)\right\} = k^n f(k) \quad (44)$$

where $f(k) = \mathbf{F}\{f(x)\}$ is the Fourier transform of $f(x)$. While on the right-hand side we use the convolution theorem.

### S2a. Using a cosine series expansion

Since we know ahead of time that the band structure of a q-BIC in an unperturbed nonlocal metasurface is periodic in momentum space, and is even by reciprocity, another natural choice for expanding the resonant frequency is the cosine series:

$$\Omega(k) = \sum_{n=0}^{\infty} \frac{b_n}{x_n^{2n}} \cos(x_n k) + i\Gamma, \quad (45)$$

where here we keep the loss constant for simplicity, $b_n$ are coefficients and

$$x_n = n\frac{P}{2\pi} \qquad (46)$$

for a lattice periodicity $P$. Then, after Fourier transformation as in the previous section, we now obtain

$$\frac{da(x,t)}{dt} = -i(i\Gamma)a(x,t) - \frac{i}{2}\sum_{n=0}^{\infty} b_n\left[a(x-x_n,t) + a(x+x_n,t)\right] + \int dx' \langle \kappa(x,x') | s_+(x') \rangle \qquad (47)$$

Now, keeping only the leading order terms $n=0,1$, we find

$$\frac{da(x,t)}{dt} = -i(i\Gamma)a(x,t) - \frac{i}{2}\left\{2b_0 a(x,t) + \frac{b_1}{x_1^2}\left[a(x-x_1,t) + a(x+x_1,t)\right]\right\} + \int dx' \langle \kappa(x,x') | s_+(x') \rangle \qquad (48)$$

Noticing that a well-known approximation for a second derivative is

$$\frac{d^2 a(x,t)}{dx^2} \approx \frac{a(x-x_1,t) + a(x+x_1,t) - 2a(x,t)}{x_1^2}, \qquad (49)$$

which becomes an equality in the limit that $x_1 \to \infty$, we re-obtain the non-Hermitian Schrodinger equation used in the main text with the equivalents

$$\begin{aligned}\omega_0 &= b_0 + \frac{b_1}{x_1^2} \\ b &= -\frac{b_1}{2}\end{aligned}. \qquad (50)$$

Given the Taylor series expansion of the cosine function, this result is well within expectation.

## S3. Comparison to bare interface

At the band-edge frequency, $\omega = \omega_0$, the nonlocal kernel drops to a factor $e^{-1}$ of its maximum value at a distance equal to

$$\xi_0 = \sqrt{b\tau_r} . \qquad (51)$$

This value, which we call the nonlocality length, is the characteristic distance across which the optical response is correlated for the band-edge mode. In general, if this value is sufficiently small, optical power is well-localized, and the metasurface may be considered local. For comparison, we consider scattering of s-polarized light from free-space to a bare interface of a dielectric with refractive index $n$, described by the Fresnel coefficient

$$r_s(k) = \frac{\sqrt{k_0^2 - k^2} - \sqrt{n^2 k_0^2 - k^2}}{\sqrt{k_0^2 - k^2} + \sqrt{n^2 k_0^2 - k^2}} . \qquad (52)$$

The nonlocal reflection kernel may be computed by taking the Fourier Transform of Eqn. (52) (setting $r_s$ to 0 when $|k| > k_0$). For comparison, Fig. S2(a,b) reproduce the Green's Function of the device from the main text, also shown in Fig. S1. Figures S2(c,d) compare the reflectance at the band-edge and magnitude of the nonlocal reflection kernel at the band to the case of a bare interface with $n = 1.45$ (glass). While there are nonzero correlations even for the bare interface (demonstrating that no optical interface is truly local), it is apparent that the $\xi_0$ for the bare interface is smaller than the wavelength, while it is larger than the wavelength in the q-BIC case. This suggests a natural definition for nonlocality: a metasurface is nonlocal if its nonlocality length is larger the wavelength in the surrounding materials.

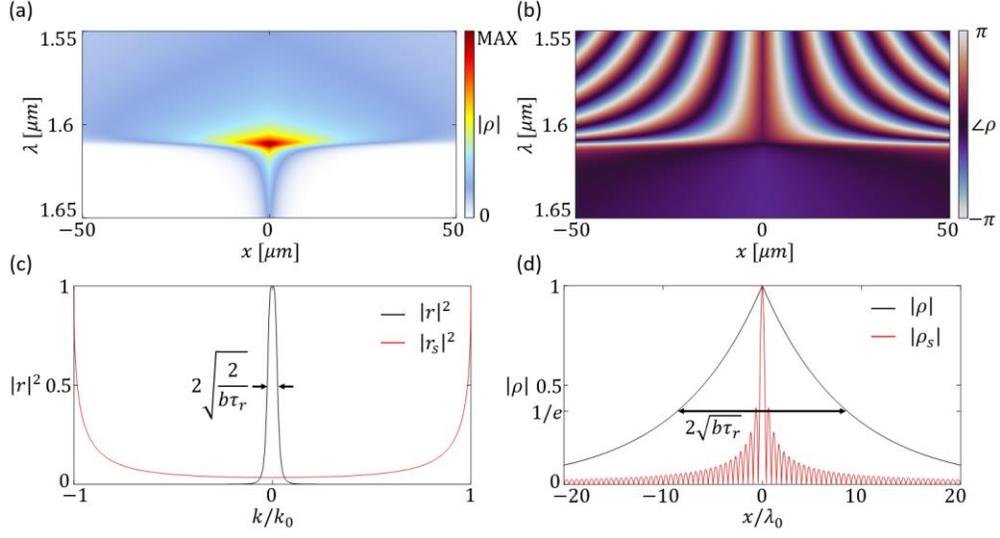

**Figure S2**. Nonlocal reflection kernel. Amplitude (a) and phase (b) of the nonlocal reflection kernel as a function of $x$ and $\lambda$ for the device in Fig. S1; copied from the main text Figure 2. (c) Comparison of the reflection coefficient $r$ for the device in Fig. 2 and the Fresnel reflection coefficient $r_s$ for s-polarized light incident on glass. (d) Comparison of the amplitude of the nonlocal kernel for the cases in (c).

## S4. Physical constraints

Here we show the derivation that the usual TCMT constraints,

$$\langle d | d \rangle = 2\gamma \tag{53}$$

$$|\kappa\rangle = |d\rangle \tag{54}$$

$$C|d^*\rangle = -|d\rangle \tag{55}$$

hold locally:

$$\langle d(x) | d(x) \rangle = 2\gamma \tag{56}$$

$$|\kappa(x)\rangle = |d(x)\rangle \tag{57}$$

$$C(x)|d^*(x)\rangle = -|d(x)\rangle \qquad (58)$$

## S4a. Equation (56)

We begin with Eqn. (56), which is derived using conservation of energy. In the case without incoming field, i.e., $|s_+\rangle = 0$, the dynamical equation (7) in the main text becomes the Schrodinger equation (11) in the main text:

$$i\frac{da(x,t)}{dt} = -\frac{b}{2}\frac{d^2 a(x,t)}{dx^2} + (\omega_0 + i\gamma)a(x,t) \qquad (59)$$

and the out-coupled wave is given by

$$|s_-(x,t)\rangle = a(x,t)|d(x,t)\rangle \qquad (60)$$

which carries a scattered power following

$$\langle s_-(x,t)|s_-(x,t)\rangle = |a(x,t)|^2 \langle d(x)|d(x)\rangle. \qquad (61)$$

Conservation of energy for such a PDE takes the form of a *continuity equation*

$$\frac{d}{dt}|a|^2 + \nabla \cdot \mathbf{j} = -\langle s_-(x,t)|s_-(x,t)\rangle \qquad (62)$$

where the "current" is given by

$$\mathbf{j} = \frac{b}{2i}\left(a^* \frac{da}{dx} - a\frac{da^*}{dx}\right). \qquad (63)$$

where we leave the $(x,t)$ dependence implicit for brevity. Note: this is consistent with conventional quantum mechanics with the replacement $\hbar=1$ and $b=1/m$.

By the chain rule, the first time in Eqn. (62) can be written

$$\frac{d}{dt}|a|^2 = a^*\frac{da}{dt} + a\frac{da^*}{dt}. \tag{64}$$

Using Eqn. (59) and its complex conjugate, we obtain

$$\frac{d}{dt}|a|^2 = -\frac{ib}{2}a\frac{d^2 a^*}{dx^2} + \frac{ib}{2}a^*\frac{d^2 a}{dx^2} + 2\gamma|a|^2. \tag{65}$$

Insertion into Eqn. (62) gives

$$2\gamma|a(x,t)|^2 = -\langle s_-(x,t)|s_-(x,t)\rangle \tag{66}$$

Which, in conjunction with Eqn. (61) finally obtains

$$2\gamma = \langle d(x)|d(x)\rangle. \tag{67}$$

**S4b. Equation (57)**

Next, we obtain Eqn. (25) using time-reversal invariance and Eqn. (24). We again consider the case without an incident source, but this time consider the time evolution of the mode subject to the initial condition: $a(x,t=0)=a_0(x)$. The modal amplitude at time $t$ is given by the propagator to be

$$a(x,t) = \int dx' a_0(x') K(x,x',t) \tag{68}$$

where

$$K(x,x',t) = \frac{1}{\sqrt{2\pi i b t}} \exp\left[i\frac{(x-x')^2}{2bt}\right] \exp[-i\omega_0 t - \gamma t], \tag{69}$$

and which decays to an outgoing wave

$$|s_-(x,t)\rangle = |d(x)\rangle \int dx' a_0(x) K(x,x',t). \tag{70}$$

By time-reversal symmetry, the time-reversed excitation of this decay process should re-obtain $a_0^*(x)$ as $t \to 0$. That is, we consider the modal amplitude given excitation by the incoming wave $|s_-^*(x,t)\rangle$ in the dynamical equation (7) of the main text, solved by the Green's Function. In particular, as $t \to 0$ we require

$$a_0^*(x) = \int dx' G_t^*(x,x',t) \langle \kappa^*(x') | s_-^*(x',t) \rangle \tag{71}$$

Inserting Eqn. (70) and writing the Green's Function in terms of the Propagator gives

$$a_0^*(x) = \frac{1}{2\gamma} \int dx' K^*(x,x',t) \langle k^*(x') | d^*(x') \rangle \int dx'' a_0^*(x'') K^*(x',x'',t) \tag{72}$$

Using the well-known behavior of Eqn. (69) that $K^*(x,x',t) \to \delta(x-x')$ as $t \to 0$, we straightforwardly arrive at

$$a_0^*(x) = \frac{1}{2\gamma} a_0^*(x) \langle \kappa^*(x) | d^*(x) \rangle \tag{73}$$

Or,

$$2\gamma = \langle \kappa^*(x) | d^*(x) \rangle \tag{74}$$

Comparison to Eqn. (67) then gives

$$|\kappa(x)\rangle = |d(x)\rangle. \tag{75}$$

**S4c. Equation (58)**

Finally, we derive Eqn. (26) of the main text. In the same scenario as the previous section, since in the forward-time case there was no source, the time reversed case should also satisfy the condition that no outgoing (scattered) waves are generated as $t \to 0$. That is, as $t \to 0$

$$0 = C(x)|s_-^*(x,t)\rangle + |d(x)\rangle \int dx' G_t(x,x',t) \langle d^*(x') | s_-^*(x,t) \rangle \tag{76}$$

Again inserting (70) and rearranging we obtain

$$C(x)|d^*(x,t)\rangle \int dx' a_0^*(x') K^*(x,x',t) = -|d(x)\rangle \frac{1}{2\gamma} \int dx' K^*(x,x',t) \langle d^*(x') | d^*(x') \rangle \int dx'' a_0^*(x'') K^*(x',x'',t)$$

$$\tag{77}$$

which, as $t \to 0$ and using Eqn. (67) becomes

$$C(x)|d^*(x)\rangle a_0^*(x) = -|d(x)\rangle a_0^*(x) \tag{78}$$

or

$$C(x)|d^*(x)\rangle = -|d(x)\rangle. \tag{79}$$

**S5. Space-frequency description of aperiodic nonlocal metasurfaces**

Regardless of periodicity, a nonlocal metasurface device may still be studied in the basis of plane waves (especially useful for phase gradient devices). A complete description of the scattering from such a metasurface requires knowledge of its scattering matrix $S(k,k',\omega)$, constructed as

$$S(k,k',\omega) = \begin{bmatrix} S_{11}(k,k',\omega) & S_{12}(k,k',\omega) \\ S_{21}(k,k',\omega) & S_{22}(k,k',\omega) \end{bmatrix}, \quad (80)$$

where $k'$ and $k$ are the basis wavevectors of the incoming and outgoing wavevectors, respectively, and subscripts delineate the two sides of the metasurface interface. For a lossless, reciprocal, system we have by reciprocity,

$$S_{ij}(k_1,k_2,\omega) = S_{ji}(-k_2,-k_1,\omega) \quad (81)$$

and by conservation of energy,

$$\begin{aligned} \int dk_1 \left[ |S_{11}(k_1,k_2,\omega)|^2 + |S_{21}(k_1,k_2,\omega)|^2 \right] = 1 \\ \int dk_1 \left[ |S_{22}(k_1,k_2,\omega)|^2 + |S_{12}(k_1,k_2,\omega)|^2 \right] = 1 \end{aligned}. \quad (82)$$

In words, in a lossless system the sum of the reflectance and transmittance to all output momenta $k_1$ must be unity for any input momentum $k_2$. Note that while in principle $k$ and $k'$ vary continuously from $-nk_0$ to $nk_0$ for surrounding media of refractive index $n$, making the dimension of these matrices infinite, in practice discretization will provide sufficient numerical accuracy and will yield a finite scattering matrix in Eqn. (80).

Now we relate the scattering matrix elements to the appropriate nonlocal kernels. By definition, the scattering matrix transforms an incoming field $E_j(k',\omega)$ into an outgoing field $E_i(k,\omega)$,

$$E_i(k,\omega) = S_{ij}(k,k',\omega) E_j(k',\omega), \qquad (83)$$

In the space frequency domain, we correspondingly have

$$E_i(x,\omega) = \int dx' \sigma_{ij}(x,x',\omega) E_j(x',\omega). \qquad (84)$$

We now seek the relationship between $S_{ij}$ and $\sigma_{ij}$. Since we wish to compare this to the scattering matrix elements in a plane wave basis, we use the test incident plane wave

$$E_j(x',\omega) = \exp(ik'x'), \qquad (85)$$

in which case Eqn. (84) shows, for instance, the reflected field is simply the Fourier transform of the nonlocal reflection kernel. We may then decompose the reflected field into its constituent plane waves by an inverse Fourier transform

$$E_i(k,\omega) = \int dx E_i(x,\omega) \exp(-ikx) \qquad (86)$$

which finally gives the reflection coefficient as

$$S_{ij}(k,k',\omega) = \int dx \int dx' \sigma_{ij}(x,x',\omega) \exp(ik'x') \exp(-ikx). \qquad (87)$$

That is to say, the elements of the scattering matrix may be computed by a mixed Fourier transform of the nonlocal kernels, confirming that the nonlocal kernels fully contain the information required to describe a nonlocal metasurface in the space-frequency domain. Likewise, inverting this relation allows retrieval of the nonlocal kernel given a known scattering matrix.

## S6. Phase gradient on and off the band edge frequency

Figure S3 studies four distinct cases: no phase gradient, at the band edge [Fig. S3(a-d)]; $k_G = 2\pi/W$, at the band edge [Fig. S3(e-h)]; no phase gradient, off the band edge [Fig. S3(i-l)]; and $k_G = 2\pi/W$, off the band edge [Fig. S3(m-p)]. Figure S3(a) shows schematically the nature of the resonant reflection at the band-edge frequency, $\omega = \omega_0$, depicting specular reflection at normal incidence when incident from either side. Figure S3 (b) reports the reflectance from side 1 in this case, showing the parabolic band structure of the q-BIC. Figure S3(c) depicts the amplitude and phase of the nonlocal reflection and transmission matrices from side 1, P and T (note that T is essentially identical to P except along the main diagonal). Finally, Fig. S3(d) reports the scattering matrix elements for this case, showing that specular transmission occurs except at normal incident from either side, wherein specular reflection occurs. Note that for sufficient resolution in k-space, Fig. S3(d) is computed using mixed fast Fourier transforms of a version of Fig. S3(c) including 10 periods.

In contrast, with a nonlocal phase gradient the device retroreflects light at the resonant frequency, as shown in Fig. S3(e). From side 1, the band edge mode is shifted in k-space by $k_G$, meaning that the resonance only occurs for $k' = k_G$. In particular,

$$\omega(k') = \omega_0 + \frac{b}{2}(k' - k_G)^2. \tag{88}$$

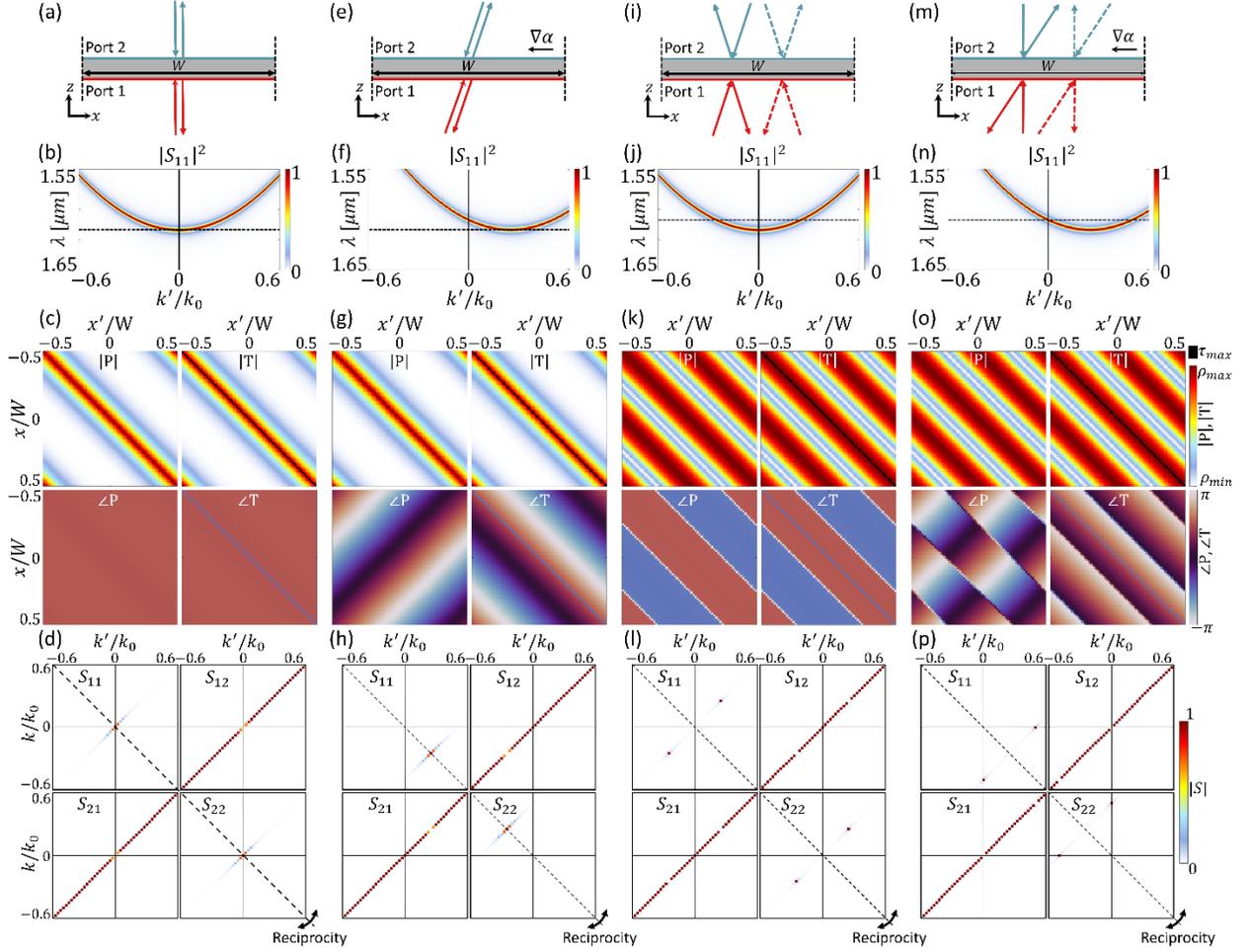

**Figure S3**. Analysis of phase gradient devices. (a) Schematic of the resonant response of a nonlocal phase gradient device with $k_G = 0$ at the band-edge frequency. (b) Reflectance as a function of incident momentum $k'$ for the device in (a). (c) Nonlocal kernel matrices for the device in (a) at the band-edge frequency. (d) Scattering matrix for the device in (a), showing unity specular reflectance only at normal incidence. (e-h) The same study in (a-d) but for a device with $|k_G| = 2\pi/W$ at the band-edge frequency. (i-l) The same study in (a-d) but at a frequency off the band edge, $\omega_1 = \omega_0 + bk_G^2/2$. (m-p) The same study in (a-d) but for a device with $|k_G| = 2\pi/W$ at a frequency off the band edge, $\omega_1 = \omega_0 + bk_G^2/2$.

This shift, as well as the anomalous reflection, is encoded in P and T in the phase information [Fig. S3(g)], and the scattering matrix at the band-edge frequency is depicted in Fig. S3(h), showing that the resonant mode occurs at $k' = k_G$ and $k = -k_G$ when light is incident from side 1, while it occurs at $k' = -k_G$ and $k = k_G$ when light is incident from side 2. Note that this behavior is consistent with the requirements of reciprocity: the scattering matrices must be symmetric about the main diagonal, which corresponds to $k = -k'$, or retroreflection. Hence we see that the action of the phase gradient is to the shift the response along the retroreflection condition by an amount $(k_G, -k_G)$ (in $S_{11}$).

Next, we consider the same device in Fig. S3(a-d) but at a frequency $\omega_1 = \omega_0 + bk_G^2/2$, which is the resonant frequency at normal incidence for the phase gradient case [see Eqn. (47)]. In this case, there are two resonant modes existing at $k' = \pm k_G$. As depicted in Fig. S3(i), when $k' = k_G$, light reflects to $k = k_G$, and under time-reversal [dashed arrows in Fig. S3(i)] we thereby naturally have the case that $k' = -k_G$ reflects to $k = -k_G$. The presence of the two modes is captured by the nonlocal kernels in Fig. S3(k), wherein we see the characteristics of a standing wave due to the two counterpropagating modes existing $k' = \pm k_G$. We also note that because we are operating off the band-edge (where Bragg scattering is maximal) the locality of the mode is substantially reduced. Upon mixed fast Fourier transformation [Fig. 4(l)], the scattering matrices confirm that these two modes correspond to two specular reflections, $k = k' = \pm k_G$, identical when light is incident from both sides (consistent with vertical symmetry).

In contrast, at the same frequency the phase gradient anomalously reflects light at normal incidence to some off-normal angle [Fig. S3(m)]. Naturally, under time-reversal this must correspond to a resonant reflection coming from the anomalously reflected angle and reflecting back to normal incidence [dashed lines in Fig. S3(m)]. As in the case without the phase gradient, these dynamics are due to the presence of two counter-propagating modes.

Here, however, instead of being excited at $k' = \pm k_G$, the phase gradient means they are excited at $k' = k_G \pm k_G$ from side 1 and $k' = -k_G \pm k_G$ from side 2. Again, these interesting dynamics are encoded in the nonlocal kernels [Fig. S3(o)], where the characteristics of counter-propagating modes are apparent but modified by a Bloch wave vector. The resulting scattering matrices confirm the picture in Fig. S3(m): the anomalous reflection angle is equal and opposite for normally incident light from side 1 versus side 2, and each of these events have a reciprocal copy (i.e., born of symmetry about the main diagonal) matching the dashed lines in Fig. S3(m).

**S7. Matrix form on a discrete grid**

Next, we show that the continuous form of the STCMT equations may be translated to discrete, or matrix form. This is done for convenience of numerical computation, and is in keeping with the tradition of metasurfaces as being built from discrete subwavelength building blocks (meta-units).

Consider, for instance, the scalar case wherein the modal parameters are invariant. For such a metasurface having meta-units at positions

$$\mathbf{x} = \begin{bmatrix} x_1 & x_2 & \ldots & x_n \end{bmatrix}^T, \tag{89}$$

we are interested in the response to incident electric fields of the form

$$\mathbf{E}(\omega) = \begin{bmatrix} E_1(\omega) & E_2(\omega) & \ldots & E_n(\omega) \end{bmatrix}^T, \tag{90}$$

where $E_i(\omega) = E(x_i, \omega)$, after interacting with a nonlocal metasurface described by the discrete phase profile

$$\mathbf{\Phi} = \pm \begin{bmatrix} \Phi_1 & \Phi_2 & \ldots & \Phi_n \end{bmatrix}^T, \tag{91}$$

where $\Phi_i = \Phi(x_i)$ and the positive (negative) sign applies to light coming from port 1 (2). For instance, in Sec. III.C the phase functions followed $\Phi(x) = \pm 2\alpha_1(x)$. Throughout we assume that the $n$ meta-units are equally spaced such that $x_i - x_{i-1} = a$. The reflected and transmitted fields may be calculated by

$$\begin{aligned} \mathbf{E}_r(\omega) &= \mathrm{P}(\omega)\mathbf{E}_{in}(\omega) \\ \mathbf{E}_t(\omega) &= \mathrm{T}(\omega)\mathbf{E}_{in}(\omega) \end{aligned}, \tag{92}$$

where

$$\begin{aligned} \mathrm{P}(\omega) &= i\frac{a}{b\tau_r}\xi(\omega)\exp\left(-\frac{|X-X'|}{\xi(\omega)}\right)e^{i\Phi_x}e^{i\Phi_{x'}} \\ \mathrm{T}(\omega) &= -iI + \frac{a}{b\tau_r}\xi(\omega)\exp\left(-\frac{|X-X'|}{\xi(\omega)}\right)e^{-i\Phi_x}e^{i\Phi_{x'}} \end{aligned} \tag{93}$$

are the nonlocal reflection and transmission kernels in matrix form (we note that the matrix forms are denoted by capitalized $\rho$ and $\tau$). Here, $I$ is the $n \times n$ identity matrix, the factor $a$ is included for normalization purposes (the choice to place it here is purely for convenience), and the $n \times n$ position matrices are given by

$$\begin{aligned} X &= \begin{bmatrix} \mathbf{x} & \mathbf{x} & \ldots & \mathbf{x} \end{bmatrix} \\ X' &= X^T \end{aligned}, \tag{94}$$

where $X$ has $n$ copies of $\mathbf{x}$. Hence, along a row the input position $x'$ varies and along a column the output position $x$ varies. Last, the phase matrices $\Phi_x$ and $\Phi_{x'}$ are populated in the same fashion as Eqn. (94):

$$\begin{aligned}\Phi_x &= \begin{bmatrix} \Phi & \Phi & \ldots & \Phi \end{bmatrix} \\ \Phi_{x'} &= \Phi_x^T \end{aligned}. \tag{95}$$

Then, the output fields due to an excitation $\mathbf{E}_{in}$ are obtained by

$$\begin{aligned} \mathbf{E}_r(\omega) &= \mathrm{P}(\omega)\mathbf{E}_{in}(\omega) \\ \mathbf{E}_t(\omega) &= \mathrm{T}(\omega)\mathbf{E}_{in}(\omega) \end{aligned} \tag{96}$$

while the reflectance and transmittance are then calculated simply as

$$\begin{aligned} R(\omega) &= \frac{\|\mathrm{P}(\omega)\mathbf{E}_{in}(\omega)\|^2}{\|\mathbf{E}_{in}(\omega)\|^2} \\ T(\omega) &= \frac{\|\mathrm{T}(\omega)\mathbf{E}_{in}(\omega)\|^2}{\|\mathbf{E}_{in}(\omega)\|^2} \end{aligned}, \tag{97}$$

where $\|\mathbf{E}\|$ is the norm of the vector $\mathbf{E}$.

Using the matrix form, we may also determine the eigen-wave of a discrete nonlocal metasurface by the eigenvector with maximal eigenvalue from the eigenvalue problem

$$R_{eig}\mathbf{E}_{eig}(\omega) = \mathrm{P}^\dagger(\omega)\mathrm{P}(\omega)\mathbf{E}_{eig}(\omega), \tag{98}$$

where † refers to the Hermitian conjugate. In full there will be *n* eigenvalues and eigenvectors; however, we are primarily interested in the eigenvector with the highest eigenvalue, which is the principal eigen-wave.

Finally, we note that the discretization introduces artificial spatial frequencies of magnitudes $m2\pi/a$, where *m* is an integer. This numerical approximation produces artificial resonances at frequencies satisfying

$$\omega = \omega_0 + \frac{b}{2}\left(m\frac{2\pi}{a}\right)^2. \tag{99}$$

Therefore, the period $a$ should be sufficiently small so as to shift these spurious modes out of the frequency range $\Delta\omega$ of interest, which will be the case if

$$a < 2\pi\sqrt{\frac{|b|}{2\Delta\omega}}. \tag{100}$$

Note the dependence on $b$: as the mode becomes more localized, a finer discretization is naturally required. The period $a$ must also be small enough to sufficiently sample the maximum phase gradient of the device, as usual for metasurfaces:

$$a < \frac{1}{N_p}\frac{\partial \Phi}{\partial x} = \frac{1}{N_p}\frac{2\pi}{P_0}, \tag{101}$$

where $N_p \geq 3$ is the desired number of phase points to be sampled as phase evolves across $2\pi$ within a distance $P_0$.

### S8. Boundary conditions for finite and infinite metasurfaces

The results of the previous section are in practice incomplete: we must take care to specify the boundary conditions. We are interested here in two boundary conditions: (i) radiative boundaries and (ii) periodic (Bloch) boundaries. We will consider each in turn. (We note that reflective boundaries may also be of interest, for instance to model guided mode resonance gratings placed between distributed Bragg reflectors [5].)

In case (i), we wish to study devices of finite size. For a finite metasurface, the energy in a q-BIC at position $x'$ is not correlated with positions $x$ that exist outside the metasurface. Hence the value of $\rho$ or $\tau$ must vanish accordingly. Happily, this boundary condition happens naturally following the procedure of the previous section. For instance, we consider the matrix form $\mathrm{P}$ at the band-edge frequency for the device in Fig. S3 but having a finite width $W = 100 \mu m$. As depicted in Fig. S4(a), this device is truncated in-plane, having radiative boundaries. The absolute value of $\mathrm{P}$ is depicted in Fig. S4(b) [while the argument of $\mathrm{P}$ is shown in Fig. S4(c)], showing peak correlation along the main diagonal and dropping exponentially off the main diagonal. Note that the main diagonal are the only populated entries in an ideally local device (see the identity factor in $\mathrm{T}$). At the upper left and lower right corners of the matrix we can see the boundary condition: locations outside the matrix implicitly encode a zero value. Considered another way, we could extend the matrix by including values of $x$ and $x'$ outside the depicted range, and populate the matrix along the diagonal according to the local reflection (which is $0$). This 'zero-padded' version would naturally be identical. Finally, Fig. S4(d) depicts the principle eigen-wave for this device, normalized to its peak value. While in the infinite case we expect a planewave, in the finite case we see that the eigen-wave's amplitude drops towards the boundaries, apparently minimizing the radiative losses. Still, the eigen-wave has a non-zero value near the boundaries, and this non-ideality results in the reflectance not being unity: the eigenvalue in this case is $R_{eig} = 0.977$. Naturally, as $W$ increase the eigen-wave better approximates a planewave, and $R_{eig}$ approaches unity.

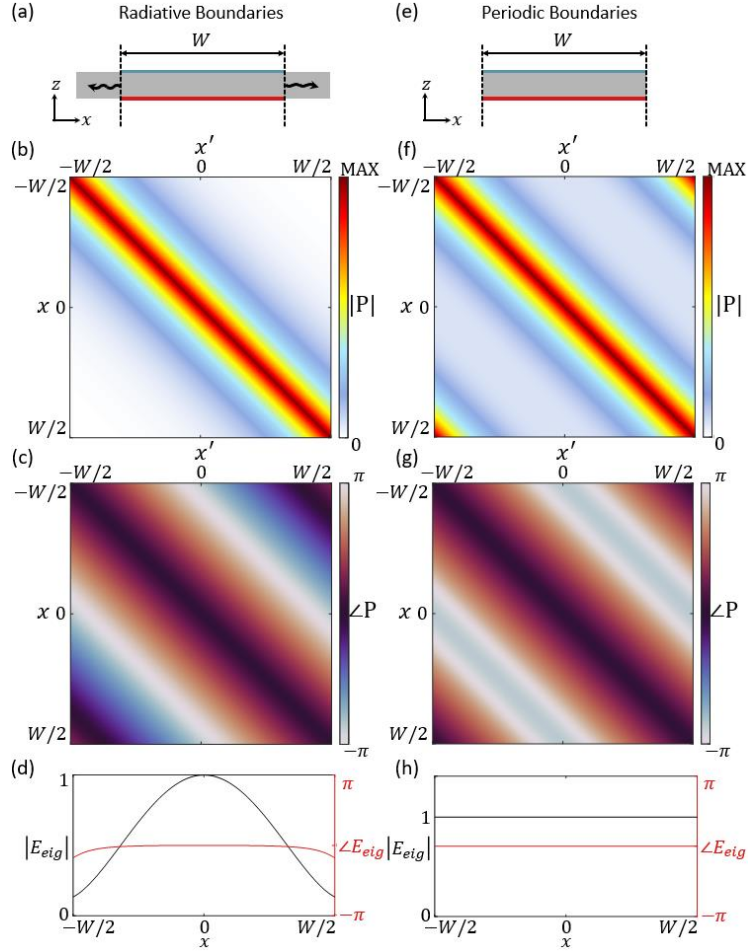

**Figure S4.** Boundary conditions for finite nonlocal kernel matrices. (a) Schematic showing radiative boundaries, wherein power in the q-BIC is not correlated with positions outside the metasurface width $W = 50\mu m$. (b,c) Amplitude and phase of the nonlocal kernel matrix $P$ with radiative boundaries for a finite version of the device in Fig. 2. (d) Principle eigen-wave for the device in (b,c). (e) Schematic showing periodic boundaries, wherein power in the q-BIC is correlated periodically with period $W$. (f,g) Amplitude and phase of the nonlocal kernel matrix $P$ for a periodic case of the device in Fig. 2. (h) Principle eigen-wave for the device in (f,g).

In case (ii), we wish to capture infinite devices, in which case we should expect the deviations from a plane wave should vanish. In the periodic case we require that the energy be correlated according to contributions from every period of the device. For instance, the left edge of the period (near $-W/2$) must be closely correlated to the right edge of the period (near $W/2$), which is not captured in the above procedure or the matrix in Fig. S4(b). Instead, we must alter the construction of P in order to include contributions from every period. For instance, for normally incident light, we sum instances of the nonlocal term shifted by integer multiples of the period $W$ of the device:

$$\begin{aligned} P(\omega) &= \sum_{j=-N}^{N} i \frac{a}{b\tau_r} \xi(\omega) \exp\left(-\frac{|X-X'-jW|}{\xi(\omega)}\right) e^{i\Phi_x} e^{i\Phi_{x'}} \\ T(\omega) &= -iI + \sum_{j=-N}^{N} \frac{a}{b\tau_r} \xi(\omega) \exp\left(-\frac{|X-X'-jW|}{\xi(\omega)}\right) e^{-i\Phi_x} e^{i\Phi_{x'}} \end{aligned} \quad (102)$$

where $2N+1$ is the number of periods accounted for. (Note that $\Phi_x = \Phi_{x-jW}$ by the periodicity of the device.) Naturally, as $N$ approaches infinity the contributions of every period in the infinite device is included, in which case the series converges to give

$$\begin{aligned} P(\omega) &= i \frac{a}{b\tau_r} \xi(\omega) \operatorname{csch}\left(\frac{W}{2\xi(\omega)}\right) \cosh\left(\frac{W-2|X-X'|}{2\xi(\omega)}\right) e^{i\Phi_x} e^{i\Phi_{x'}} \\ T(\omega) &= -iI + \frac{a}{b\tau_r} \xi(\omega) \operatorname{csch}\left(\frac{W}{2\xi(\omega)}\right) \cosh\left(\frac{W-2|X-X'|}{2\xi(\omega)}\right) e^{-i\Phi_x} e^{i\Phi_{x'}} \end{aligned} \quad (103)$$

Figure 4(e) schematically depicts a periodic device with period $W = 100\,\mu m$, and the resulting matrix P is shown in Figs. S4(f,g). The boundary conditions are apparent here especially in the upper right and lower left corners of Fig. S4(f), where we see that the main diagonal appears to 'wrap', accounting for the periodicity. Finally, Fig. S4(h) shows the normalized eigen-wave for this case, which is a plane wave with eigenvalue is $R_{eig} = 1$, as expected.

Lastly, since the boundary conditions must be consistent with both the device and the illumination, for light incident with momentum $k'$ we include the Bloch wave in each period of the summation:

$$P(k',\omega) = i \sum_{j=-N}^{N} \frac{a}{b\tau_r} \xi(\omega) \exp\left(-\frac{|X - X' - jW|}{\xi(\omega)}\right) e^{i\Phi_x} e^{i\Phi_{x'}} e^{ik'W}$$
$$T(k',\omega) = -iI + \sum_{j=-N}^{N} \frac{a}{b\tau_r} \xi(\omega) \exp\left(-\frac{|X - X' - jW|}{\xi(\omega)}\right) e^{-i\Phi_x} e^{i\Phi_{x'}} e^{ik'W} \quad (104)$$

However, the mixed Fourier transform gives the scattering matrices as a function of both $k'$ and $k$, allowing us to simply use the case without the Bloch wave seen in Eqns. (103).

## S9. 2D nonlocal metasurfaces

In this section, we briefly demonstrate the extension of the STCMT to two-dimensional (2D) metasurfaces. This required extending the nonlocal kernels to be a function of both $x$ and $y$: $\rho(x,x',y,y',\omega)$ and $\tau(x,x',y,y',\omega)$, making the corresponding discrete matrix forms four-dimensional. However, we can flatten the 4D structure into a 2D array of matrices for ease of use and visualization. The 2D array is $N_y \times N_y$ in dimension, where $N_y$ is the number of $y$ positions in the metasurface. Rows of the array correspond to changing $y'$, while columns correspond to changing $y$. Each element is then a $N_x \times N_x$ matrix as in the above sections, where $N_x$ is the number of $x$ positions.

As an example, Fig. S4(a) depicts the reflection matrix for the case of a radial nonlocal metalens, discretized from

$$\rho(x,x',y,y',\omega) = \frac{i}{b\tau_r} \xi(\omega) \exp\left(-\frac{\sqrt{(x-x')^2 + (y-y')^2}}{\xi(\omega)}\right) e^{i\Phi(x,y)} e^{i\Phi(x',y')}, \tag{105}$$

with

$$\Phi(x,y) = -k_0 \sqrt{x^2 + y^2 + f^2}. \tag{106}$$

For simplicity, we assume that the band structure is completely isotropic. Figure S4(b) shows the central few elements of the array in Fig. S4(a), overlaid with the indexing of this array described above. Figures S4(c,d) similarly show the phase of the reflection matrix. Each element of the array resembles the cylindrical lenses studied in the previous section, but the radial phase of function of the device is recovered by reshaping the main diagonal of Fig. S4(c) onto the corresponding $(x,y)$ grid, depicted in Fig. S4(e). As usual, we expect the characteristic response to be encapsulated by the eigen-wave of this device. Figures S4(f,g) depict the amplitude and phase of the metalens, where a similar reshaping is done to the eigenvector for visualization purposes. Finally, Fig. S4(h) depicts the refocused spot at $z = -f$ after excitation by the eigen-wave, consistent with expectation.

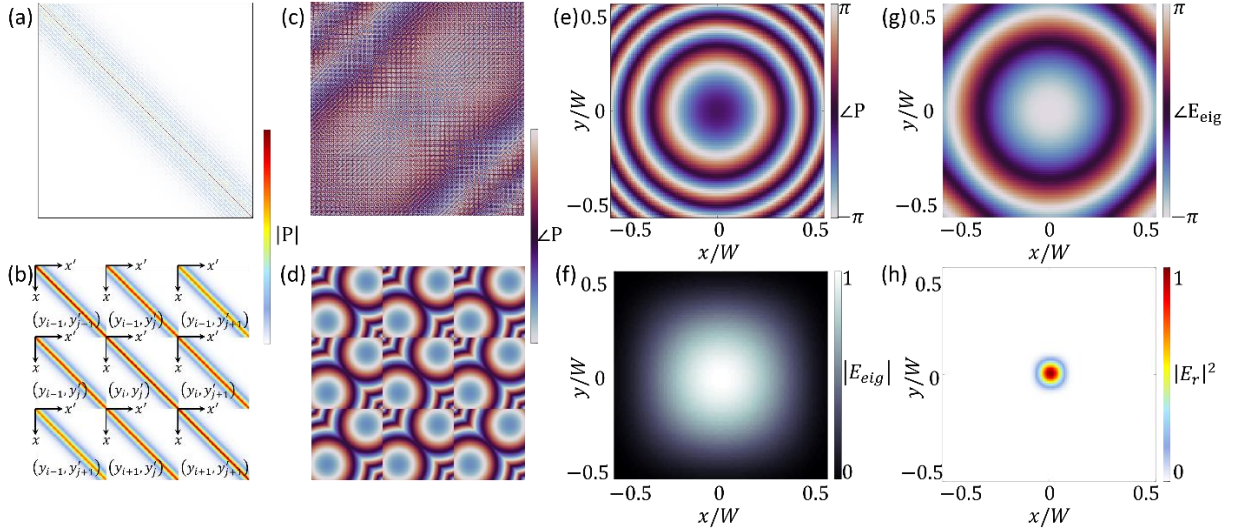

**Figure S5**. 2D nonlocal metalens. (a,c) Amplitude and phase of the nonlocal reflection kernel, a 2D array of matrices. (b,d) Zoom-in of the central regions of (a,c), overlaid with indexing of the array of matrices. (e) Main diagonal of (c), reshaped onto an $(x,y)$ grid to depict the characteristic phase function of a metalens. (f,g) Amplitude and phase of the computed eigenwave, which focuses upon reflection at $z = -f_0$. (h) Intensity of the reflected light at the focal plane.

Lastly, we note that the matrix size (which has important consequences for memory requirements and other computational costs) scale as $N^4$, where $N$ is the characteristic number of positions in each $x$ and $y$. This compares poorly to full-wave simulations, in which the field matrices scale as $N^2$ for fixed $z$ dimension simulations. However, it is apparent from Fig. S5 that the vast majority of elements are $0$, suggesting the use of sparse matrices. In this case, we recover the $N^2$ scaling once the dimension $W$ of the device is larger than the nonlocality length: increasing the number of positions only effectively extends the region near the main diagonal, while all the other elements remain 0.

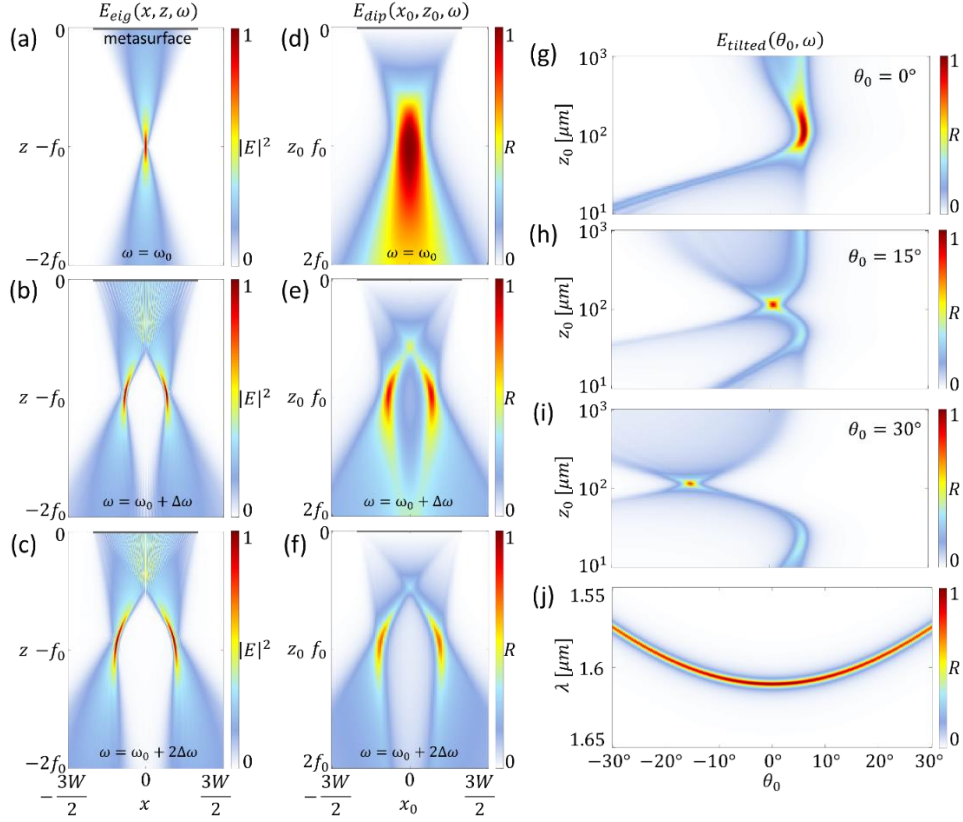

**Figure S6**. Eigen-waves and spatial selectivity, on and off the band-edge. (a) Eigen-wave for a metalens at $\omega=\omega_0$, propagating downwards after reflection from the metasurface at $z=0$. (b,c) Eigen-waves as in (a) but at frequencies $\omega=\omega_0+\Delta\omega$ and $\omega=\omega_0+2\Delta\omega$, where $\Delta\omega=0.01\mu m^{-1}$. Note the comatic focal responses. (d) Reflectance due to an ideal point source as a function of the position of the source, $(x_0,z_0)$. (e,f) Response as in (d) but at frequencies $\omega=\omega_0+\Delta\omega$ and s$\omega=\omega_0+2\Delta\omega$. (g) Reflectance due to a point source placed along the optical axis ($x_0=0$). (h,i) Reflectance due to a point source placed along the optical axis but for a modification of a tilt factor $\exp[ink_0\sin(\theta_0)]$ with tilt angles $\theta_0$. (j) Reflectance due to tilted point sources placed at $(0,-f_0)$ as a function of $\theta_0$, recovering the band structure of the q-BIC.

## S10. Nonlocal metalenses off the band edge frequency

Here, we extend our study of the nonlocal metalenses to frequencies off the band edge. We begin by computing the principle eigen-wave at three frequencies and studying how they propagate. Figure S6(a-c) shows the intensity of the reflected eigen-wave at the band edge [Fig. S6(a)], a frequency $\omega_0 + \Delta\omega$ [Fig. S6(b)] and $\omega_0 + 2\Delta\omega$ [Fig. S6(c)], where $\Delta\omega = 0.01 \mu m^{-1}$. The band-edge eigen-wave refocuses to a point $z = -f$, as expected, but at the two other frequencies, we see that this focal spot as split into two comatic focuses off the optical axis, where the split is larger for the case that is further off the band edge. In parallel, we compute the reflectance as a function of $(x_0, z_0)$ at the same three frequencies, and comparing the results [Figs. S6(d-f)] shows that the strong relationship between the eigen-wave and spatial selectivity extends off the optical axis and off the band edge. We also see that the selectivity increases drastically off the band edge, consistent with the fact that the nonlocality length increases off the band edge (due to reduced Bragg scattering).

To understand this behavior further, we study the reflectance spectra due to a modified excitation

$$E_{in}(x,\omega) = \exp\left(ik_0\sqrt{x^2 + z_0^2}\right)\exp\left[ik_0 \sin(\theta_0)\right], \tag{107}$$

which represents a point source with a momentum shift (e.g., as if it were deflected by a local metasurface before impinging on the nonlocal metasurface). Figures S6(g-j) depict the spectral reflectance for three values of $\theta_0$, showing that the peak reflectance stays at $z_0 = f$ but blue shifts with larger $\theta_0$. This suggests that the modes off the band edge are involved; indeed, Fig. S6(k) depicts the reflectance when $z_0 = f$ as a function of $\theta_0$, recovering the underlying band structure of the q-BIC.

From these results, we see that the q-BIC off the band edge is selective for waves having the characteristics of the eigen-wave at the band edge, but modified by the momentum of the off-band-edge modes. Yet when a single mode is involved, as discussed in the previous section, the reflected wave is not a time-reversed copy. Instead, just as the reciprocal copies of reflection events in the phase gradient case (see Fig. S3), we have two reciprocal focal spots: when a point source excites the nonlocal metalens at one of the foci in Fig. 4(e) of the main text, for instance, the reflected light refocuses to the other focus. Hence, a superposition of light originating from both foci will refocus light back to both foci; this is the eigen-wave of the device at this off-band-edge frequency.

**S11. Wavefront-shaping vs wavefront-selective regimes**

A plane wave at normal incidence excites a resonant frequency depending on the local value of the phase gradient [as in Eqn. (47)]:

$$\omega_r(x) = \omega_0 + \frac{b}{2}\left(\frac{\partial \Phi}{\partial x}\right)^2. \tag{108}$$

If the range of frequencies as a function of position is larger than the linewidth of the resonance, the reflectance will naturally be low, meaning we have a wavefront-selective device. In a metalens, the gradient varies between a value of 0 (at the center of the device) to a value $k_0\text{NA}$. Hence the range of resonant frequencies is $\Delta\omega_r = \frac{|b|}{2}(k_0\text{NA})^2$. Enforcing $\Delta\omega_r < d\omega$, where $d\omega = \omega_0/Q$ is the linewidth of the resonance, we arrive at the condition for the wavefront-shaping regime:

$$\text{NA} < \sqrt{\frac{2/k_0}{|b|Q}}. \tag{109}$$

Notably, the nonlocality length of the band-edge mode is $\xi_0 = \text{Re}[\xi(\omega_0)] = \sqrt{b\tau_r} = \sqrt{bQ/\omega_0}$, which is the characteristic distance that the q-BIC travels in-plane before coupling out. We may rewrite Eqn. (54) in terms of the nonlocality length as

$$\text{NA} < \frac{1}{\sqrt{2}\pi} \frac{\lambda_0}{\xi_0}. \tag{110}$$

## S12. Spatial selectivity and eigen-waves

Here, we comment on the expanded understanding of the eigen-wave in these systems. The rigorous definition provided here makes precise the underlying physics demonstrated in Ref. [2] and allowed extension of the concept to frequencies off the band edge and to eigen-waves with submaximal reflectance. We highlight two fundamental features gleaned in the resulting study: (i) there is a strong relationship between the eigen-waves and the point response of the system and (ii) the eigen-waves (regardless of frequency) are best understood as the superposition of two reciprocal reflection events. Regarding (i), the eigen-wave is not only the wave that, when incident, reflects with maximal efficiency to its time-reversed copy, it is also highly characteristic of the spatial selectivity of the device. We may understand this further by considering eigenvectors other than the one with highest reflectance (i.e., what we call the 'principle' eigen-wave). As shown in Fig. S7, which depicts the eigenvalues for a nonlocal metalens at the band-edge frequency (sorted in descending order of magnitude), the reflectance of the eigenvectors drops quickly away from the principle eigen-wave. Considering the eigendecomposition of the matrix (i.e., $P^*P = U\Lambda U^{-1}$, where $U$ is the matrix composed of the eigenvectors and $\Lambda$ is the diagonal matrix whose diagonal contain the corresponding eigenvalues), we may also consider the reflectance of an arbitrary wave by the appropriate sum of the inner products of the eigenvectors with the incident wave, weighted by the appropriate eigenvalues. With this perspective and in light of Fig. S7, it is immediately clear why the eigen-wave characterizes the spatial selectivity: only incident waves with large inner products (i.e., high degree of match) with the principle eigen-wave have substantial reflectance.

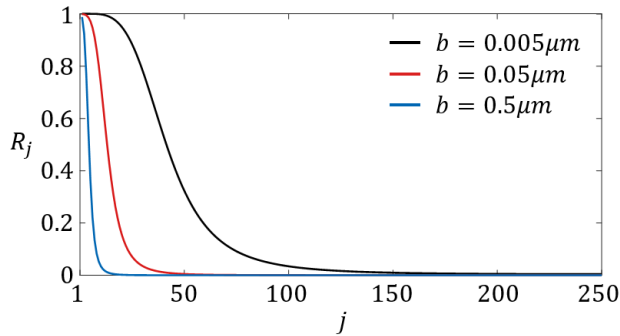

**Figure S7.** Sorted eigenvalues $R_j$ for a nonlocal metalens with three different values of $b$, showing increased selectivity (i.e., an effectively smaller basis) to incoming waves as $b$ increases.

### S13. Thermal metasurfaces off the band-edge frequency

Here, we apply STCMT to study thermal metalenses as a function of frequency. Figure S8 depicts the computed response for three metalenses with varying Q-factors, $Q=10^2$, $Q=10^3$, and $Q=10^4$. Figures S8(a,c,e) show the response in the space-frequency domain, showing the response roughly follow a parabolic condition ($b=0.15\mu m$ here) while demonstrating increasing selectivity to the position of a point source as the Q-factor increases, especially off the band-edge. Also shown is the selectivity to the spin of the point source. Figures S89(b,d,f) show spatial absorption maps at example wavelengths of interest, showing the splitting of the focal spot into two, as is Fig. S6. Interestingly, as the Q-factor grows, the off-band-edge absorption diminishes rapidly. This implies having a high Q-factor not only decreases the FWHM at the band-edge frequency, but also limits the range of frequencies away from the band-edge contributing meaningful absorption.

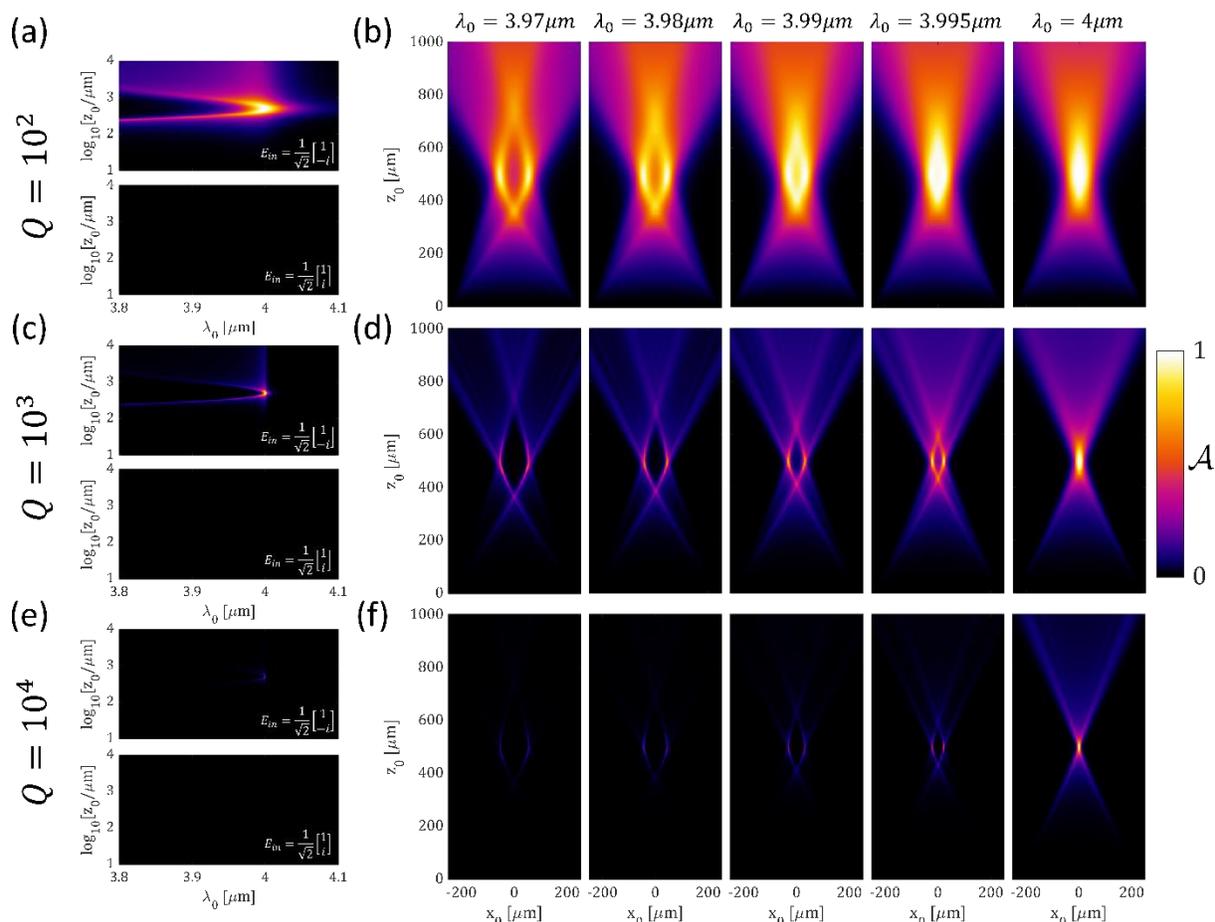

**Figure S8.** Thermal metalenses as a function of frequency. (a,c,e) Absorption as a function free-space wavelength $\lambda_0$ and position of a point source placed at $(0, z_0)$ for the two spin states. (b,d,f) Absorption for the selected spin state as a function of point sources placed at $(x_0, z_0)$ for select wavelengths.

### S14. Applicability to local and nonlocal metasurface design

We identify a few key lessons applicable to the broader field of metasurfaces as follows. (i) The nonlocal kernels contain all the information of the functionality of the device, described in the space-frequency domain; (ii) the scattering matrix defined by input and output momenta contains this same information, related to the nonlocal kernels by a mixed Fourier transform; (iii) the nonlocal kernels of an aperiodic device may be approximately constructed by reference to a library of nonlocal kernels of individual structures; and (iv) the scattering matrix of the composite device is then retrievable by a suitable mixed Fourier transformation.

Additionally, we identify the lessons particular to aperiodic nonlocal metasurfaces as: (i) the nonlocal phase gradient requires vertical asymmetry in the case of unity efficiency, imparting equal and opposite momentum from above or below; (ii) the eigen-waves of the device are composed of a superposition of two waves that reflect into each upon interaction with a nonlocal metasurface; (iii) the eigen-waves closely characterize the point response function of the device, and thereby the spatial selectivity; (iv) the degree of nonlocality increases with lifetime and decreases as the band becomes flatter; (v) the spatial selectivity depends on the degree of nonlocality and the range of momenta encoded into the device; and (vi) a nonlocal metasurface may be classified as either wavefront-shaping or wavefront-selective.

**S15. Limitations of the present study**

Here, we discuss some limitations in these initial studies. First, we note that our analytical study, and indeed all the devices in Refs. [2], was limited to when the local reflection coefficient is near 0. This choice follows the finding in Refs. [3],[4] that this condition yields complete control over the q-BIC scattering phase. Future work is needed to go beyond this condition and will be aided by the insights of the STCMT. Second, we note that the parabolic approximation of the band structure fails when the spatial frequencies become too high, implying either more terms are necessary in the Taylor expansion and likely requiring purely numerical approaches. Third, we assumed throughout that the eigenpolarization is independent of incident angle; yet studies on BICs have demonstrated polarization vortices in the momentum-frequency domain [8]. These systems may be captured by modeling the polarization dependence and including its variance in the Fourier transforms. Fourth, we assumed that the local Fresnel coefficient was independent of incident angle (i.e., the reflection coefficient is 0 for all spatial frequencies); yet it is well known that all interfaces have unity reflectance at grazing incidence. The consequence of this assumption is the absence of highly localizes features such as those seen in Fig. S1 for a bare interface, and this simplification yielded the convenient and insightful analytical relations we found. However, if the nonlocality length is comparable or smaller than the characteristic extent of these features from the local scattering, and if we operate near grazing incidence, our approximation is invalid, and we must return to numerical methods. Finally, we remind the reader of the spurious numerical results born of sparse discretization, implying that the bandwidth of the STCMT is not unlimited and increasing it requires increasing the resolution and therefore size of the nonlocal matrices.

**Supplementary References**